\documentclass[a4paper,12pt,authoryear]{article}
\usepackage{xcolor,amsmath}
\usepackage{amsthm,bm}
\usepackage{hyperref}
\usepackage{amssymb}
\usepackage{dsfont}
\usepackage{comment}
\usepackage{float}
\usepackage{graphicx}
\usepackage{pbox}
\usepackage{colortbl}
\usepackage{epstopdf}
\usepackage{mathtools}
\usepackage{multirow}
\usepackage{relsize}
\usepackage{bigints}
\usepackage{cases}
\usepackage[font={footnotesize}]{caption}
\usepackage{hhline}
\usepackage{booktabs}
\usepackage{bbm}
\usepackage{array}
\usepackage{algorithm2e}
\usepackage{adjustbox}
\usepackage{longtable}
\usepackage{ltablex}
\usepackage{caption}
\usepackage{subcaption}

\usepackage{natbib}

\AtBeginDocument{}

\theoremstyle{remark}

\theoremstyle{plain}

\newcommand*\diff{\mathop{}\!\mathrm{d}}

\title{A Loss-Based Prior for Gaussian Graphical Models}

\usepackage{authblk}

\author[1]{Laurentiu Catalin Hinoveanu\thanks{lch36@kent.ac.uk}}
\author[1]{Fabrizio Leisen\thanks{fabrizio.leisen@gmail.com}}
\author[2]{Cristiano Villa\thanks{cristiano.villa@ncl.ac.uk}}
\affil[1]{School of Mathematics, Statistics and Actuarial Science, University of Kent}
\affil[2]{School of Mathematics, Statistics and Physics, Newcastle University}

\usepackage[english]{isodate}

\usepackage{parskip}
\date{}

\begin{document}
\maketitle

\begin{abstract}
\noindent
Gaussian graphical models play an important role in various areas such as genetics, finance, statistical physics and others. They are a powerful modelling tool which allows one to describe the relationships among the variables of interest. From the Bayesian perspective, there are two sources of randomness: one is related to the multivariate distribution and the quantities that may parametrise the model, the other has to do with the underlying graph, $G$, equivalent to describing the conditional independence structure of the model under consideration. In this paper, we propose a prior on G based on two loss components. One considers the loss in information one would incur in selecting the wrong graph, while the second penalises for large number of edges, favouring sparsity. We illustrate the prior on simulated data and on real datasets, and compare the results with other priors on $G$ used in the literature. Moreover, we present a default choice of the prior as well as discuss how it can be calibrated so as to reflect available prior information.
\end{abstract}

\section{Introduction}\label{sc_introduction}

A graphical model can provide a geometrical representation of the dependencies among variables with the immediacy that graphs exhibit. The use of this particular type of models is widespread within disciplines, including finance and economics \citep{Giudici2016}, social sciences \citep{McNally2015,Williams2018}, speech recognition \citep{Bilmes2004,Bell2007} and biology \citep{Wang2016}.

A widely used statistical model for graphs is the \textit{Gaussian Graphical Model} (GGM). There are many useful reasons for assuming Normality. A remarkable one is that, among all distributions with same mean and same variance, the Normal assumption maximizes the entropy, see for example \cite{Cover2006}. As a consequence, it  imposes the least number of structural constraints beyond the first
and second moments. As such, the focus of this paper is on GGM. 

We are tackling the Gaussian graphical model problem from the Bayesian perspective. In this approach there are two source of randomness as discussed by \citet{Giudici}. One is related to the multivariate distribution and the quantities that may parametrise it, the other has to do with the underlying graph $G$, equivalent to describing the conditional independence structure of the model under consideration. As such, two kinds of priors are necessary: one related to the model parameters, $\Sigma_G$ in our case, the other associated with the graph $G$. In this paper, we will focus on assigning a loss-based prior on $G$, through the methodology of \citet{VillaModel}. 

The idea is to consider the loss that one would incur in not choosing a particular graph, when this graph is the true one. In other words, we perform a thought experiment in objectively measuring both the loss in information and the loss due to complexity derived by selecting the ``wrong'' graph. The loss in information is determined by the minimum expected Kullback--Leibler divergence from graph, say $G$, and the graph, say $G^\prime$; that is, the graph that it is the most similar to $G$. The loss due to complexity will consists of two components: an \emph{absolute} one, proportional to the number of edges of the graph, and a \emph{relative} one, proportional to the number of graphs with the same size as $G$. As detailed in Section \ref{sc_theprior}, the loss in information will generally be zero, as such the prior on $G$ will be:

\begin{equation*}
\pi^{(h,c)}(G) \propto \exp\left\lbrace-h\left[(1-c)|G|+c\log\binom{m}{|G|}\right]\right\rbrace,\qquad h>0,\quad 0\leq c\leq 1,
\end{equation*} 
where $m$ represents the maximum number of possible edges in graph $G$.

We will show in Section \ref{sc_theprior} that the above prior encompasses some proposals currently used in the literature. For example, the one discussed in \cite{Carvalho} or the uniform prior.

The paper has the following structure. In Section 2 we provide some preliminaries on Gaussian graphical models and some of the graph priors used in this framework. Section 3 introduces our proposed graph prior. We outline the behaviour of our prior for simulated and real data examples in Section 4. Section 5 contains some final discussion points.

\section{Graph priors for Gaussian graphical models}\label{sc_methodology}

In this section we will set the scene to introduce our prior. The  first subsection provides some background on Gaussian graphical models. The second section illustrates the state of the art of graph priors in the objective framework. However, the following subsections require some preliminaries. 

Following \citet{Lauritzen:1996}, a graph $G$ is represented by the pair $G=(V,E)$ with $V$ a finite set of vertices and $E$ a subset of $V\times V$ of pairs of distinct edges. Throughtout the paper we will consider 
$V=\{1, 2, \ldots ,p\}$, where $p$ is a strictly positive integer. A subset of $V$ is an $(i,j)-$separator when all the paths from $i$ to $j$ go through the respective subset. Subset $C\subseteq V$ separates $A$ from $B$ if  $C$ is a $(i,j)-$\textit{separator} $\forall \: i \in A, j \in B$. A graph where $(i,j)\in E, \forall \: i,j \in V$ is called a \textit{complete} graph. A subgraph represents a subset of $V$ such that the edge set is restricted to those edges that have both endpoints in the respective subset. We call a maximal complete subgraph a \textit{clique}. We refer to the decomposition of an undirected graph as a triple $(A,B,C)$ where $V=A\cup C \cup B$ for disjoint sets $A,C$ and $B$ such that $C$ separates $A$ from $B$ and $C$ is complete. Therefore, the graph is decomposed in the subgraphs $G_{A\cup C}$ and $G_{B\cup C}$. A decomposable graph can be broken up into cliques and separators.  For a non-decomposable graph there will be subgraphs which cannot be decomposed further and are not complete. 

\subsection{Gaussian graphical models}

Let $\mathbf{X}=(X_1, X_2, \ldots, X_p)^\mathbf{T}$ be a $p$-dimensional random vector which follows a multivariate Gaussian distribution, that is
\begin{equation}
\mathbf{X} \sim N_p(\bm{0}, \Sigma_G)\nonumber,
\end{equation}
where $\bm{0}\in \mathbb{R}^{p}$ is a $p$-dimensional column vector of zero means and $\Sigma_G \in \mathbb{R}^{p\times p}$ is the positive-definite covariance matrix. Let $\mathbf{x}=(\mathbf{x}_1, \mathbf{x}_2,  \ldots, \mathbf{x}_n)^{\mathbf{T}}$ be the $n\times p$ matrix of observations, where $\mathbf{x}_i$, for $i=1,\ldots,n$, is a $p$-dimensional realisation from the multivariate Gaussian distribution. The link between the assumed sampling distribution and the graph is specified by completing a positive definite matrix with respect to an undirected graph \citep{Roverato1998,  Giudici, Massam}. For an arbitrary positive definite matrix $\Gamma$ and an undirected graph $G$, $\Sigma_G$ is the unique positive definite matrix completion of $\Gamma$ with respect to $G$. This means that for the pairs of vertices which share an edge, the corresponding entries of $\Sigma_G$ are the same as $\Gamma$. The entries for the missing edges are set to be $0$ in the concentration (precision) matrix, that is $\Sigma_G^{-1}$. Therefore, we have a link between the multivariate sampling distribution and the graph structure represented by the zeros of the concentration matrix $\Sigma_G^{-1}$. In the Gaussian graphical models framework, the dimension $p$ of the multivariate Gaussian distribution also represents the number of vertices in the undirected graph $G$. As our sampling distribution is Gaussian, the concentration matrix has a clear interpretation. The entries of the concentration matrix encode the conditional independence structure of the distribution \citep{Lauritzen:1996}. As such, if and only if the $(i,j)^{\textrm{th}}$ element of the concentration matrix is 0, the random variables $X_i$ and $X_j$ are conditionally independent given all other variables in the matrix (pairwise Markov property); or, equivalently, given their neighbours (local Markov property). The previous statement is based upon the idea that in a Gaussian graphical model the global, local and pairwise Markov properties are equivalent. 

\medskip

The literature around Gaussian graphical models is vast, and it spans from  frequentist to Bayesian approaches. \citet{Meinhausen2006} estimate the neighbourhood of vertices through the LASSO procedure \citep{Tibshirani1996} and then combine those estimates to build the underlying graph. Of the same flavour as LASSO, \citet{Yuan} have introduced a penalized likelihood method to estimate the concentration matrix, which for Gaussian graphical models encodes the conditional independence. \cite{Friedman2008} have developed the graphical LASSO algorithm which is quite fast compared to other frequentist based algorithms. The above methods look at the regularization penalty being imposed on the concentration matrix. A method where the penalty is imposed to the inverse of the concentration matrix, the covariance matrix, is presented by \citet{Bien2011}. \citet{Giudici} have applied the trans-dimensional reversible jump Markov chain Monte Carlo (RJMCMC) algorithm of  \citet{Green1995} to estimate the decomposable graphs that underlie the relationships in the data. This RJMCMC method was extended to estimate the structure in a case of multivariate lattice data by \citet{DobraRJMCMC}. Another trans-dimensional algorithm, this time based upon birth-death processes, was described by \citet{Mohammadi2015}. \citet{jones2005} have reviewed the traditional MCMC (Markov chain Monte Carlo) methods used for graph search for both decomposable and non-decomposable cases when high-dimensional data is considered and have proposed an alternative method to find high probability regions of the graph space. An MCMC method to estimate the normalising constant of the distribution which has its structure characterised by a non-decomposable graph has been proposed by \citet{Massam}. Their idea was also used by \citet{jones2005} when non-decomposable graphs were involved. For decomposable graphs, \citet{Carvalho} have introduced a prior for the covariance matrix which helps to improve the accuracy in the graph search. In addition, they have also presented a graph prior which automatically guards against multiplicity.

The estimation methods in GGMs have been extensively studied in the literature for both directed (\citet{Friedman2000}, \citet{Spirtes2000}, \citet{Geiger2002}, \citet{Shojaie2010}, \citet{Stingo2010}, \citet{Yajima2015}, \citet{Consonni2017}) and undirected graphs (\cite{Dobra2004}, \cite{Meinhausen2006}, \cite{Yuan}, \cite{Banerjee2008}, \cite{Friedman2008}, \cite{Carvalho}, \cite{Kundu2019}, \cite{Stingo2015}).

\subsection{Graph priors}

Assuming $G$ decomposable, \citet{Giudici}  discuss the following prior on $G$:
\begin{equation}
\pi(G)=d^{-1}\nonumber,
\end{equation}
where $d$ is the number of decomposable graphs on a specific vertex set $V$.
If we consider unrestricted graphs, the above prior is the uniform prior on the graph space and  has the form:
\begin{equation}
\pi(G)=\dfrac{1}{2^{\binom{|V|}{2}}}.
\label{uniform_prior}
\end{equation}
where $|V|$ is  the number of vertices in the graph.  A criticism in using a uniform prior is that it assigns more mass to medium size graphs compared to, for example, the empty graph or the full graph. 

To address the problem, \citet{jones2005} set independent Bernoulli trials on the edge inclusions, such that the prior probability is $\phi=2/(|V|-1)$ leading to an expected number of edges equal to $|V|$. Thus, the prior on $G$ is:
\begin{equation}
\pi(G)\propto \phi^{|G|}\cdot(1-\phi)^{m-|G|}\nonumber,
\end{equation}
where  $0\leq |G| \leq m$ is the number of edges in the graph $G$ and $m=\binom{|V|}{2}$ represents the maximum number of possible edges in that respective graph. Clearly, a $\phi$ close to zero would encourage sparser graphs, while for $\phi\rightarrow 1$, more mass will be put on complex graphs.

\citet{Carvalho} recommend a fully Bayesian approach, where $\phi$ should be inferred from the data. As such, they assume that $\phi\sim \textrm{Beta}(a,b)$, leading to:
\begin{equation}
\pi(G)\propto \dfrac{\beta\left(a+|G|,b+m-|G|\right)}{\beta(a,b)}.
\label{Carvalho}
\end{equation}
By setting $a=b=1$ (equivalent to setting a uniform prior on $\phi$) in equation \eqref{Carvalho}, they obtain the prior on $G$ as: 
\begin{equation}
\pi(G)\propto \dfrac{1}{\left(m+1\right)}\dbinom{m}{|G|}^{-1}.
\label{Carvalho_Beta_1_1}
\end{equation}
A property of the prior in equation \eqref{Carvalho_Beta_1_1} is that it corrects for multiplicity. That is,  as more noise vertices are added to the true graph, the number of false positives (edges which are erroneously included in the graph) remains constant.

A somewhat similar form of the prior in equation \eqref{Carvalho_Beta_1_1} was derived by \citet{Armstrong2009}. Their prior, called the \textit{sized based prior}, uses the $A_{p,|G|}$ parameter representing the number of decomposable graphs instead of the combinatorial coefficient in the  formula from above. The value of $A_{p,|G|}$ is estimated using an MCMC scheme and a recurrence relationship with graphs that have up to 5 vertices.  

\section{A loss-based prior for Gaussian graphical models}\label{sc_theprior}
In this section, we present a prior based on a methodology that involves loss functions, firstly introduced in \citet{VillaModel}.

To outline their idea, let us consider $k$ Bayesian models
\begin{equation*}
M_{j}=\{f_{j}(x|\theta_{j}), \pi_{j}(\theta_{j})\}\qquad   j\in\{1,2,\ldots, k\},
\label{modeldef}
\end{equation*}
where $f_{j}(x|\theta_{j})$ is the sampling distribution parametrised by $\theta_{j}$ and  $\pi_{j}(\theta_{j})$ represents the prior on the model parameter (possibly vector of parameters) $\theta_{j}$. 
Assuming the priors $\pi_j(\theta_j)$ are proper, the model prior probability $P(M_j)$ is proportional to the expected Kullback--Liebler divergence from $M_j$ to $M_i$, with $i=1,\ldots,k$ and $i\neq j$, where the expectation is considered with respect to  $\pi_{j}(\theta_{j})$. That is: 
\begin{align}
P(M_{j})\propto &\exp\left\lbrace  \mathbb{E}_{\pi_{j}}\left[\inf_{\theta_{i},i\neq j} D_{KL}(f_{j}(x|\theta_{j})\|f_{i}(x|\theta_{i}))\right]\right\rbrace, \qquad    j=1, \ldots, k. \label{VillaModelCompactOrg}
\end{align}
In other words, we assign a prior mass to model $M_j$ which is proportional to the distance to the most similar model $M_i$ $(i\neq j)$, in expectation. To illustrate, let us start by considering what is lost if model $M_j$ is removed from the set of all the possible models and it is the true model. This loss is quantified by the Kullback--Leibler divergence from $M_j$ to its nearest model. The loss is then linked to the model prior probability via the self-information loss function \citep{Merhav1998}. The prior in \eqref{VillaModelCompactOrg} is then obtained by equating the two above losses. The methodology has been used in the framework of change point analysis \citep{Hinoveanu2019} and  for variable selection in linear regression models \citep{VillaLee}. We follow the insight provided by the latter by adding an additional loss component to account for model complexity. We designed the penalty term to penalize complex graphs, meaning graphs with a relatively large number of edges. ~For instance, this is in line with the approach suggested by  \citet{Cowell2007}. Therefore, for a given number of vertices $p$ with a maximum number of edges $m$, our prior has the form
\begin{align}
\pi^{(h,c)}(G) &\propto \exp\left\lbrace  \underbrace{\mathbb{E}_{\pi}\left[\inf_{\Sigma_{G^\prime},} D_{KL}(f(\mathbf{x}|\bm{0},\Sigma_{G})||f(\mathbf{x}|\bm{0},\Sigma_{G^\prime}))\right]}_{\textrm{loss due to information}} \right.\nonumber \\
&\left.\underbrace{-h\left[(1-c)|G|+c\log\binom{m}{|G|}\right]}_{\textrm{loss due to graph complexity}}\right\rbrace,
\label{OurPrior}
\end{align}
with $h \in [0,+\infty)$ and $c \in [0, 1]$.
The component of the prior that penalizes for complexity takes into account the number of the edges of the graph, $|G|$, as well as the number of graphs with the same number of edges, $\binom{m}{|G|}$. The former can be interpreted as an absolute complexity of the graph, whilst the latter is weighing the  complexity of the graph relatively to all the graphs with the same number of edges (i.e. relative complexity). Note that the last one is considered in the log-scale for mitigating the exponential behaviour of the binomial coefficient for large $m$. This makes the two terms approximately on the same order of magnitude. The two components are mixed by means of $c$, while $h$ represents the constant up to which a loss function is defined.
To fully understand the behaviour of the prior, in equation \eqref{OurPrior} we first note that, if $G$ is a non-complete graph, then the loss due to information is zero. In fact, one would be able to find a graph $G'$ in which $G$ is nested; then, the minimum expected Kullback--Leibler divergence is attained for  $\Sigma_{G^\prime}=\Sigma_{G}$, therefore zero. When $G$ is the complete graph, the minimum expected Kullback--Leibler divergence will generally be larger than zero as the complete graph cannot be nested in any other graph. However, as we show in Appendix B, the minimum expected Kullback--Leibler divergence from the complete graph is negligible and we can approximate the loss in information to zero. Thus, the proposed prior becomes,
 \begin{equation}
\pi^{(h,c)}(G) \propto \exp\left\lbrace-h\left[(1-c)|G|+c\log\binom{m}{|G|}\right]\right\rbrace,
\label{RefinedOurPrior}
\end{equation} 
which is a sensible choice for any element in the space of graphs.

The constant $h$ allows to set the prior in order to control the sparsity of the graph. In particular, for $h\rightarrow\infty$, the prior in equation \eqref{RefinedOurPrior} will decrease quickly to zero, assigning most of the mass to simple graphs.
On the other hand, small values of $h$ result in a prior where is more evenly distributed over the whole space of graphs. In fact, if we set $h=0$ the prior in \eqref{RefinedOurPrior} will become $\pi(G)\propto 1$, that is the uniform prior. An interesting feature of the prior in \eqref{RefinedOurPrior} is that it has, as particular cases, other well-known priors, besides the uniform prior. By setting, $c=1$ and $h=1$ we recover the  prior in equation \eqref{Carvalho_Beta_1_1} proposed by \citet{Carvalho}. 

If we set $c=0$ we obtain 
\begin{equation}\label{eq_vlprior}
\pi^{(h,0)}(G) \propto \exp\left\lbrace -h|G|\right\rbrace, 
\end{equation}
which has a similar structure to the prior of  \citet{VillaLee}, introduced in the context of linear regression. For some particular choices of $h$ and $c$, our prior is reduced to some well-known graph priors from the relevant literature as seen in Table \ref{Table_prior_notation_specification}.

\begin{table}[H]
\centering
\begin{tabular}{c|c|c|c}
\hline 
$h$ & $c$ & Prior & Equation \\ 
\hline 
0 & $c$ & Uniform prior & \eqref{uniform_prior} \\ 
1 & 1 & \citet{Carvalho} prior & \eqref{Carvalho_Beta_1_1} \\
$h$ & 0 & \citet{VillaLee} prior  & \eqref{eq_vlprior}  \\
1 & 1/2 & Mixture prior  & \eqref{eq_mpprior}  \\
\hline 
\end{tabular}
\caption{The particular values of $h$ and $c$ that lead to our prior being reduced to some popular graph priors from the relevant literature.}
\label{Table_prior_notation_specification}
\end{table}

Let $M(G)$ represent the set of symmetric positive-definite matrices constrained by $G$, which means there is an equivalence between the zeroes of the concentration matrix $\Sigma_G^{-1}$ and the missing edges from graph $G$. The function $f(\mathbf{x}|\Sigma_G,G)$ denotes the multivariate Gaussian sampling distribution with covariance matrix $\Sigma_G$. Then, the graph posterior probability is
\begin{equation*}
\pi(G|\mathbf{x}) \propto \pi(G)\int_{\Sigma_G\in M(G)}f(\mathbf{x}|\Sigma_G,G)\pi(\Sigma_G|G)\diff \Sigma_G.
\end{equation*}

Although our prior is suitable for both decomposable and non-decomposable graphs, here we mainly focus on the former class of graphs so that we can compare the performance of our prior to other priors available in the literature.

Regarding the marginal likelihood, we are using the hyper-inverse Wishart $g$-prior of \citet{Carvalho} as the prior for the constrained covariance matrix $\Sigma_G$. This prior arises as the implied fractional prior of the covariance matrix \citep{OHagan1995} for the following noninformative prior, whose form was purposely selected to maintain conjugacy
\begin{equation*}
\pi_N(\Sigma|G)\propto \dfrac{\prod_{C\in \mathcal{C}} \det(\Sigma_C)^{-|C|}}{\prod_{S\in \mathcal{S}} \det(\Sigma_S)^{-|S|}}.
\end{equation*}
Here, $\mathcal{C}$ and $\mathcal{S}$ represent the clique and separator sets for  graph $G$, respectively.
Furthermore, the hyper-inverse Wishart $g$-prior ($0<g<1$) is a conjugate prior for the multivariate Gaussian distribution. As such, the marginal likelihood can be expressed in closed form as (see \citet{Carvalho})
\begin{equation*}
f(\mathbf{x}|G)=(2\pi)^{-np/2}\dfrac{H_G(gn,g\mathbf{x}^T\mathbf{x})}{H_G(n,\mathbf{x}^T\mathbf{x})},
\label{FractionalLikelihood}
\end{equation*}
with $H_G(b,D)$ denoting the normalising constant of the hyper-inverse Wishart distribution with degrees of freedom parameter $b \in \rm I\!R^{+}$ and scale matrix $D \in M(G)$.  For a decomposable graph, $H_G(b,D)$ can be expressed as a ratio of products over the cliques and separators (see \citet{Carvalho, Dawid1993}), that is
\begin{equation*}
H_G(b,D)=\dfrac{\prod_{C\in \mathcal{C}} \det\left(\dfrac{1}{2}D_C\right)^{\dfrac{b+|C|-1}{2}}\Gamma_{|C|}\left(\dfrac{b+|C|-1}{2}\right)^{-1}}{\prod_{S\in \mathcal{S}} \det\left(\dfrac{1}{2}D_S\right)^{\dfrac{b+|S|-1}{2}}\Gamma_{|S|}\left(\dfrac{b+|S|-1}{2}\right)^{-1}},
\end{equation*}
where 
\begin{equation*}
\Gamma_a(x)=\pi^{\dfrac{a(a-1)}{4}}\prod_{j=1}^a\Gamma(x+(1-j)/2)
\end{equation*}
represents the multivariate gamma function.

As recommended by \citet{Carvalho}, in all our further analyses we set $g=1/n$. To explore the graph space we have used the feature-inclusion stochastic search (FINCS) algorithm of \citet{Scott2008}. An outline of the algorithm is provided in the Appendix A.

\section{Simulated and Real Data Examples}\label{sc_illustrations}
In this section, we are showing the behaviour of the prior in equation \eqref{RefinedOurPrior} in both simulated and real data scenarios. We focus on decomposable graphs and inference is made by implementing the FINCS algorithm.

For the analyses, on simulated and real data, we compare four priors on $G$. Namely, the Carvalho and Scott prior ($\pi^{(1,1)}(G)$), the uniform prior ($\pi^{(0,c)}(G)$) and the proposed prior with two different settings: in the first we have $h=1$ and $c=0$ ($\pi^{(1,0)}(G)$) and for the second we have $h=1$ and $c=0.5$ ($\pi^{(1,1/2)}(G)$). Thus:
\begin{equation}
\pi^{(1,0)}(G) \propto \exp\{-|G|\}\nonumber
\end{equation}
and 
\begin{equation} \label{eq_mpprior}
\pi^{(1,1/2)}(G) \propto \exp\left\{-\left[\frac{1}{2}|G| + \frac{1}{2} \binom{m}{|G|}\right]\right\}.
\end{equation}
The above setting choices have been dictated by the following reasons. The  prior $\pi^{(1,0)}(G)$ highlights the choice of a prior that penalises for the absolute graph complexity without including any prior information on the rate of penalisation (controllable by setting $h$). The choice of the $\pi^{(1,1/2)}(G)$ prior is driven by the motivation of understanding how equal weights for the two types of penalty considered, i.e. absolute versus relative, interplay.

\subsection{Simulated Data Example}\label{sc_simulated}
\begin{figure}[H]

\includegraphics[clip, trim=0cm 0cm 0cm 2cm, scale=1]{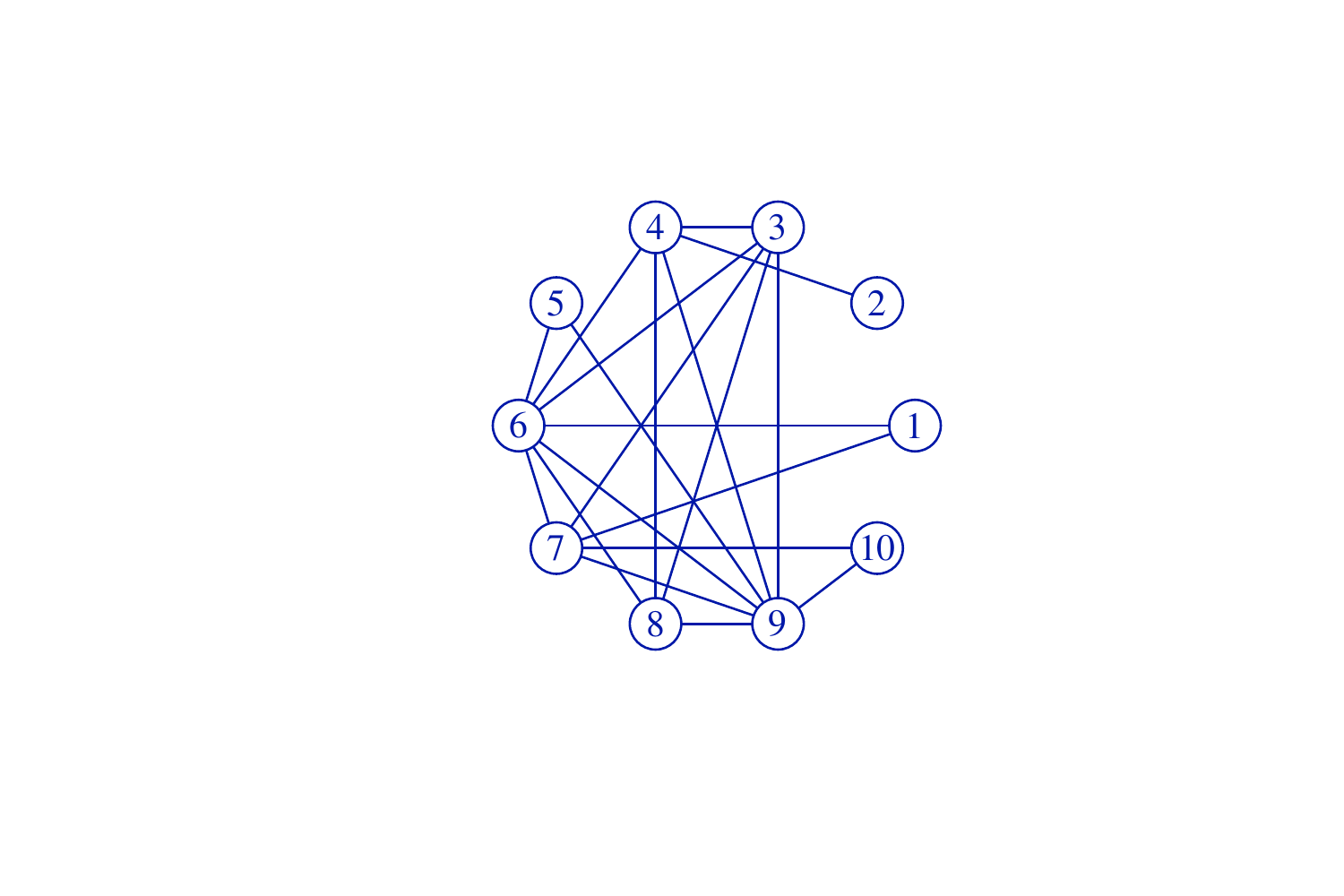}
\centering
\vspace*{-3cm}
\caption{Plot of the 10 vertices graph used in the first simulation study.}
\label{Simulated_Graph}
\end{figure}

The first simulation study has been taken from \citet{Carvalho}. We start from a graph with 10 vertices and 20 edges, which is represented in Figure \ref{Simulated_Graph}. We have then added 5 and 40 noise vertices for, respectively, the first and the second simulation. These noise vertices represent vertices unconnected to each other or with the 10 vertices graph. The data has been simulated from a zero mean multivariate normal distribution with the covariance matrix designed to represent the dependencies of the above graphs. In both cases the sample size was of $n=50$ observations. That is, we have sampled 50 realisations for a $p=15$ vertices graph and a $p=50$ vertices graph, where each graph contains just the edges shown in Figure \ref{Simulated_Graph}, through the \textsf{R} package BDgraph of \citet{Moham}. For the simulated data, we are using a single covariance matrix for each of the two cases. More precisely, to simulate the data we have used the \texttt{bdgraph.sim()} function with the following arguments: the adjacency matrix was given respectively by one of the two graph structures described previously and the $G$-Wishart prior was the default one. We have run FINCS for 5 million iterations and set a global move every 50 iterations; the resampling step was considered at every 10 iterations. During the FINCS search, we have saved the best 1000 graphs. The estimated edge posterior inclusion probabilities were computed as
\begin{equation*}
\hat{q}_{ij}=\dfrac{\sum_{r=1}^t\mathbbm{1}_{(i,j)\in G_r} f(\mathbf{x}|G_r)\pi(G_r)}{\sum_{r=1}^t f(\mathbf{x}|G_r)\pi(G_r)},
\end{equation*} 
where $t$ is the number of uniquely discovered graphs in terms of the log-score. The posterior inclusion probabilities for each of the 22 edges of the graphs are reported in Table \ref{TableResults_5Nodes} and Table \ref{TableResults_40Nodes}, for the cases $p=15$ and  $p=50$, respectively. These tables also report the number of false negatives flags (FNs) and false positives (FPs) under each prior. 

\medskip

\begin{tabularx}{\textwidth}{ccccc}
\hline 
\multirow{2}{*}{Edge} & \multicolumn{4}{c}{Noise vertices: 5 ($p$=15)}\\ 
\cline{2-5} 
& \pbox{10cm}{\relax\ifvmode\centering\fi \rule{0pt}{1em}$\pi^{(1,1)}(G)$ \rule[-7pt]{0pt}{0pt}} & \pbox{5cm}{\relax\ifvmode\centering\fi \rule{0pt}{1em}$\pi^{(1,0)}(G)$  \rule[-7pt]{0pt}{0pt}} & \pbox{5cm}{\relax\ifvmode\centering\fi \rule{0pt}{1em}$\pi^{(1,1/2)}(G)$  \rule[-7pt]{0pt}{0pt}} & \pbox{10cm}{\relax\ifvmode\centering\fi \rule{0pt}{1em}$\pi^{(0,c)}(G)$ \rule[-7pt]{0pt}{0pt}} \\ 
\hline 
(1,6) & 0.167 & 0.234  & 0.216 & 0.158 \\ 
(1,7) & 0.916 & 0.981  & 0.960 & 0.997 \\ 
(2,4) & 0.079 & 0.173  &  0.126 & 0.184 \\ 
(3,4) & 0.014 & 0.017 & 0.018 & 0.321  \\ 
(3,6) & 0.961 & 0.994 & 0.987 & 0.999 \\ 
(3,7) & 0.198 & 0.355  & 0.282 & 0.311 \\ 
(3,8) & 0.997 & 1.000 & 0.999 & 1.000 \\ 
(3,9) & 0.013 & 0.012 & 0.013 & 0.025 \\
(4,6) & 0.023 & 0.025 & 0.027 & 0.366  \\ 
(4,8) & 0.005 & 0.003 & 0.005 & 0.006 \\ 
(4,9) & 0.493 & 0.877 & 0.721 & 0.984  \\ 
(5,6) & 0.007 & 0.003 & 0.005 & 0.007  \\ 
(5,9) & 0.698 & 0.958 & 0.878 & 0.994  \\
(6,7) & 0.014 & 0.014 & 0.015 & 0.013 \\ 
(6,8) & 0.005 & 0.009 & 0.007 & 0.018  \\ 
(6,9) & 0.011 & 0.013 & 0.011 & 0.297  \\
(7,9) & 0.213 & 0.153 & 0.179 & 0.097 \\ 
(7,10)& 1.000 & 1.000 & 1.000 & 1.000 \\  
(8,9) & 0.006 & 0.007 & 0.007 & 0.015  \\ 
(9,10) & 0.785 & 0.874 & 0.834 & 0.962 \\
\hline  
FPs: & 0 & 1 & 0 & 2\\
\hline 
FNs: & 14 & 13 & 13 & 13\\
\hline
\caption{Estimated posterior inclusion probabilities for the true graph edges together with the number of  false positives (FPs) and false negatives (FNs) flags when there are 5 noise nodes and with $p=15$.}
\label{TableResults_5Nodes}
\end{tabularx}


\begin{tabularx}{\textwidth}{ccccc}
\hline 
\multirow{2}{*}{Edge} & \multicolumn{4}{c}{Noise Vertices: 40 ($p$=50)}\\ 
\cline{2-5} 
& \pbox{5cm}{\relax\ifvmode\centering\fi \rule{0pt}{1em}$\pi^{(1,1)}(G)$ \rule[-7pt]{0pt}{0pt}} & \pbox{5cm}{\relax\ifvmode\centering\fi \rule{0pt}{1em}$\pi^{(1,0)}(G)$ \rule[-7pt]{0pt}{0pt}} & \pbox{5cm}{\relax\ifvmode\centering\fi \rule{0pt}{1em}$\pi^{(1,1/2)}(G)$  \rule[-7pt]{0pt}{0pt}} & \pbox{5cm}{\relax\ifvmode\centering\fi \rule{0pt}{1em}$\pi^{(0,c)}(G)$  \rule[-7pt]{0pt}{0pt}} \\ 
\hline 
(2,4) & 0.454 & 0.996  &  0.753 & 1.000 \\ 
(3,4) & 0.002 & 0.003 & 0.003 & 0.120 \\ 
(3,6) & 0.000 & 0.000 & 0.000 & 0.000 \\ 
(3,7) & 0.000 & 0.000  & 0.000 & 0.000 \\ 
(3,9) & 0.001 & 0.001 & 0.001 & 0.006 \\
(4,6) & 0.000 & 0.000 & 0.000 & 0.000 \\ 
(4,9) & 0.089 & 0.001 & 0.016 & 0.002 \\ 
(5,6) & 0.000 & 0.000 & 0.000 & 0.001 \\ 
(6,8) & 0.000 & 0.000 & 0.000 & 0.001 \\ 
(7,10)& 0.000 & 0.000 & 0.000 & 0.001 \\  
(8,9) & 0.912 & 1.000 & 0.985 & 1.000 \\ 
\hline
\pbox{4cm}{\relax\ifvmode\centering\fi \rule{0pt}{1em} True graph's edges with posterior inclusion probability $>0.95$ under all considered priors \rule[-7pt]{0pt}{0pt}} & \multicolumn{4}{c}{\pbox{6cm}{\relax\ifvmode\centering\fi \rule{0pt}{1em}(1,6), (1,7), (3,8), (4,8), (5,9), (6,7), (6,9), (7,9), (9,10)\rule[-7pt]{0pt}{0pt}}}\\ 
\cline{2-5} 
\hline  
FPs: & 0 & 11 & 2 & 41 \\
\hline
FNs: & 10 & 9 & 9 & 9 \\
\hline
\caption{Estimated posterior inclusion probabilities for the true graph edges together with the number of  false positives (FPs) and false negatives (FNs) flags when there are 40 noise nodes and with $p=50$.}
\label{TableResults_40Nodes}
\end{tabularx}

In terms of FPs, we see an increase for the $\pi^{(1,0)}(G)$ and $\pi^{(0,c)}(G)$ priors when moving from 5 to 40 noise vertices; although of different size. In fact, the $\pi^{(1,0)}(G)$ prior moves from 1 to 11 false positives, while the $\pi^{(0,c)}(G)$ prior moves from 2 to 41. For the $\pi^{(1,1/2)}(G)$ prior, that is when we mix the $\pi^{(1,0)}(G)$ and the $\pi^{(1,1)}(G)$ prior with equal weight, the increase in FPs is marginal. In terms of FNs, we note minimal differences among the priors with an additional false positive under the Carvalho and Scott's prior in both cases. 

To compare the inferential results of the prior we consider the median probability graphs, that is the graphs composed by all the edges with a posterior inclusion probability of at least 0.5. Note that, there are other approaches to choose the threshold for posterior edge inclusion probability; see for example \cite{Newton2004}. However, to ease the comparison with the results in \cite{Carvalho}, we have set the threshold as above. In both cases the priors yield to similar graphs, with the exception of edge (4, 9) for the experiment with $p=15$ and (2, 4) for the experiment with $p=50$. The above edges are not included in the graph derived by using the $\pi^{(1,1)}(G)$ prior, although the posterior inclusion probability is close to 0.5 (0.49 and 0.45, respectively).

In the second simulation exercise, we study the performance of the proposed prior when initial information about the number of edges is available (and one wishes to reflect this in the prior). The results are compared to the ones obtained by using the Bernoulli prior implemented in the BDgraph package. We consider both the case of accurate prior information as well as the case where the prior information about the true number of edges is not accurate. We have considered the scenarios with $n=50$ and $n=100$, and we have repeated the analysis for $250$ randomly generated samples. Computational details are that we have employed the BDgraph package appropriately modified to allow the implementation of our prior, and we have run $200000$ iterations with the burn-in of $100000$. First, we have simulated from a graph with $6$ vertices and $3$ edges, and assumed that the prior information about the expected number of edges was correct. To have prior distributions with mean of $3$, we have set $h=0.28$ and $c=0.11$ for the $\pi^{(h,c)}(G)$ prior and choose a probability of success of $0.2$ for the Bernoulli prior. We have compared the two priors by considering the average size of the posterior graph over the $250$ samples. Table \ref{BDgraph_sim_not_misspecified} shows the statistics of the simulation study, including the $99\%$ bootstrap confidence interval based on one million replicates. We note that the $\pi^{(0.28,0.11)}(G)$ prior outperforms the Bernoulli prior as the confidence intervals contain the true graph size for both $n=50$ and $n=100$. For the second case, we have sampled from a graph with $6$ vertices and $5$ edges assuming that the prior information about the true graph size is as before (i.e. $3$ edges). If we keep the $\pi^{(h,c)}(G)$ prior with the same setting as above, the $99\%$ confidence intervals are $(4.04,4.56)$ and $(4.25,4.76)$ for, respectively, $n=50$ and $n=100$. However, the $\pi^{(h,c)}(G)$ prior allows to set $h$ and $c$ to have the same prior mean as above and a larger variance. In Table \ref{BDgraph_sim_misspecified} we report the frequentist summaries for $\pi^{(1.36,0.93)}(G)$ prior with a variance of $35.5$ and the Bernoulli prior. We note, in this case, that the confidence intervals for the $\pi^{(1.36,0.93)}(G)$ prior contain the true number of edges. Although there is a discrepancy between the $\pi^{(1.36,0.93)}(G)$ prior variance and the Bernoulli prior (2.4), this shows a higher versatility of the $\pi^{(h,c)}(G)$ prior as it allows to control two pieces of prior information (mean and variance) by the choice of the parameters $h$ and $c$.

\begin{table}[H]
\centering
\resizebox{\textwidth}{!}{%
\begin{tabular}{c|cc|cc}
\hline 
\multirow{2}{*}{Prior} & \multicolumn{2}{c|}{$n=50$} & \multicolumn{2}{c}{$n=100$} \\ 
& Average Size & 99\% Confidence Interval & Average Size & 99\% Confidence Interval \\ 

\hline 
$\pi^{(0.28,0.11)}(G)$ & 2.81  & (2.61, 3.02) & 2.96  & (2.78, 3.14) \\
Bernoulli & 1.88  & (1.74, 2.04) & 2.22   & (2.08,  2.36)  \\ 
\hline 
\end{tabular}%
}
\caption{Frequentist summaries for the $\pi^{(0.28,0.11)}(G)$ prior and the Bernoulli prior when prior information is accurate}
\label{BDgraph_sim_not_misspecified}
\end{table}

\begin{table}[H]
\centering
\resizebox{\textwidth}{!}{%
\begin{tabular}{c|cc|cc}
\hline 
\multirow{2}{*}{Prior} & \multicolumn{2}{c|}{$n=50$} & \multicolumn{2}{c}{$n=100$} \\ 
& Average Size & 99\% Confidence Interval & Average Size & 99\% Confidence Interval \\ 

\hline 
$\pi^{(1.36,0.93)}(G)$ & 5.47  & (4.64, 6.34) & 4.50  & (3.98, 5.08) \\
Bernoulli & 3.36  & (3.13, 3.60) & 3.82   & (3.61, 4.03)  \\ 
\hline 
\end{tabular}%
}
\caption{Frequentist summaries for the $\pi^{(1.36,0.93)}(G)$ prior and the Bernoulli prior when prior information is not accurate}
\label{BDgraph_sim_misspecified}
\end{table}

\subsection{Real Data Examples}\label{sc_realdata}
In this section we illustrate our prior in real data scenarios. We compare the performance of our prior with the other priors considered in the previous section.  We have selected three data sets, encompassing different sizes, both in terms of variables and in terms of number of observations. The results, obtained with the same settings for the FINCS algorithm implemented in Section \ref{sc_simulated}, are presented in the next subsections. For comparison purposes, edges have been selected as part of the estimated graph if their posterior inclusion probability was at least 0.5 (median probability graph).

\subsubsection{The Multivariate Flow Cytometry Dataset}\label{sc_cytometrydata}
\citet{Sachs2005} consider flow cytometry measurements for 11 phosphorylated proteins and phospholipids across a total number of 7466 observations. The 11 proteins considered have the following nomenclature: Raf, Mek, Plcg, PIP2, PIP3, Erk, Akt, PKA, PKC, P38, Jnk. The purpose of their study was to infer a Bayesian network to reveal possible connections between enzymes. We have centred the data and the key results are reported in Table \ref{Full_Sachs_common} and Table \ref{Full_Sachs_omitted}.

The most sparse graph was produced using the $\pi^{(1,0)}(G)$ prior, and the included edges are listed in Table \ref{Full_Sachs_common}. In Table \ref{Full_Sachs_omitted} , we can see the edges that were omitted for the $\pi^{(1,0)}(G)$ prior, but included for the others. The most complex graph is selected under the $\pi^{(1,1)}(G)$ prior, where 5 extra edges are added, while the $\pi^{(1,1/2)}(G)$ and the $\pi^{(0,c)}(G)$ priors include, respectively, 1 and 2 edges more than the $\pi^{(1,0)}(G)$ prior. To note, edge $(1,8)$, which is included by all the priors except the $\pi^{(1,0)}(G)$ prior, has a posterior inclusion probability for the latter prior relatively close to 0.5, suggesting that it is likely to be the sole relevant difference among the priors. For the remaining edges in Table \ref{Full_Sachs_omitted}, a more conservative threshold (e.g. set at 0.7) would have excluded them from all the graphs. For the included edges (Table \ref{Full_Sachs_common}), there is strong agreement among the priors, as the posterior inclusion probabilities are all quite close to one. The above results can also be noted in Figure \ref{Fig:FlowCyto}, where we have plotted the graph representing the edges common to all the priors, as well as the edges peculiar to each prior. 

\vspace*{1cm}
\begin{tabularx}{\textwidth}{cccccc}
  \hline
Index  & Edge & $\pi^{(1,1)}(G)$  & $\pi^{(1,0)}(G)$ & $\pi^{(1,1/2)}(G)$ & $\pi^{(0,c)}(G)$ \\ 
  \hline 
  1 & (2,10) & 0.892 & 0.932 & 0.907 & 0.904 \\ 
  2 & (3,9) & 0.978 & 0.910 & 0.952 & 0.957 \\ 
  3 & (5,11) & 0.947 & 0.938 & 0.924 & 0.923 \\ 
  \hline
\multicolumn{2}{c}{\pbox{5cm}{\relax\ifvmode\centering\fi \rule{0pt}{1em} Edges with posterior inclusion probability $>0.95$ under all considered priors \rule[-7pt]{0pt}{0pt}}} & \multicolumn{4}{c}{\pbox{8cm}{\relax\ifvmode\centering\fi \rule{0pt}{1em}(1,2), (1,3), (1,6), (1,7), (1,11), (2,3), (2,6), (2,7), (2,8), (2,11), (3,4), (3,5), (3,6), (3,7), (3,8), (3,10), (3,11), (4,5), (5,7), (6,7), (6,8), (6,11), (7,8), (7,9), (7,10), (7,11), (8,9), (8,10), (8,11), (9,10), (9,11), (10,11)\rule[-7pt]{0pt}{0pt}}}\\ 
\cline{1-6} 
\caption{Edges with a posterior inclusion probability of at least 0.5 for all four priors considered.}
\label{Full_Sachs_common}
\end{tabularx}

\begin{tabularx}{\textwidth}{cccccc}
  \hline
Index  & Edge & $\pi^{(1,1)}(G)$  & $\pi^{(1,0)}(G)$ & $\pi^{(1,1/2)}(G)$ & $\pi^{(0,c)}(G)$ \\ 
  \hline 
  1 & (1,5)  & 0.550 & 0.043 & 0.182 & 0.216 \\ 
  2 & (1,8) & 0.832 & 0.436 & 0.644 & 0.677 \\ 
  3 & (2,5) & 0.561 & 0.046 & 0.190 & 0.224 \\ 
  4 & (2,9) & 0.656 & 0.322 & 0.480 & 0.507 \\ 
  5 & (4,11) & 0.528 & 0.197 & 0.338 & 0.363 \\ 
\caption{Edges with a posterior inclusion probability smaller than 0.5 under the $\pi^{(1,0)}(G)$ prior, but with a value larger than 0.5 under at least one of the other three priors.}
\label{Full_Sachs_omitted}
\end{tabularx}


\begin{figure}[H]
  \centering
    \subcaptionbox{The number of common edges is 35}{\includegraphics[width=0.5\textwidth]{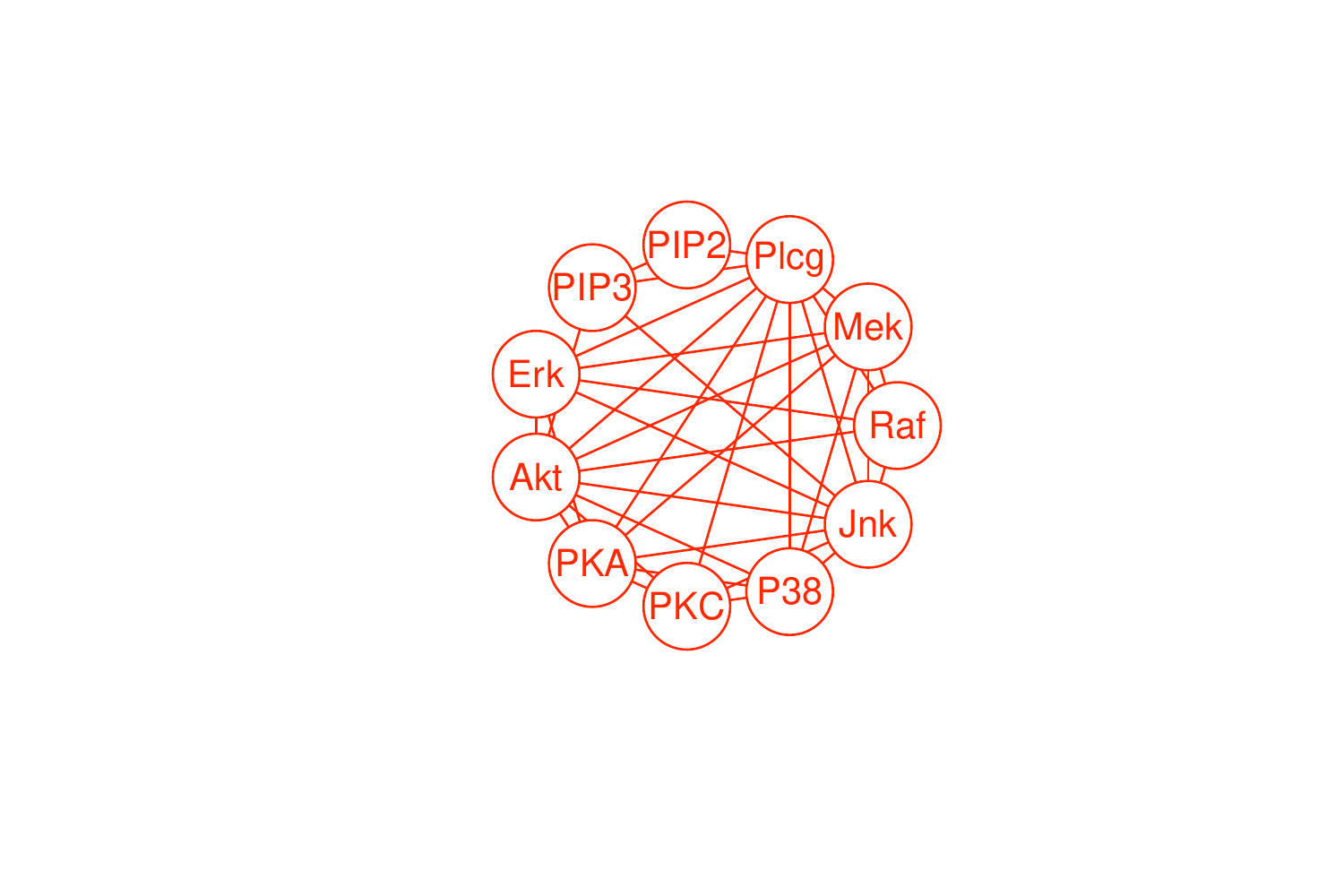}}
  \subcaptionbox{$\pi^{(1,1)}(G)$ with 5 extra edges}{\includegraphics[width=0.5\textwidth]{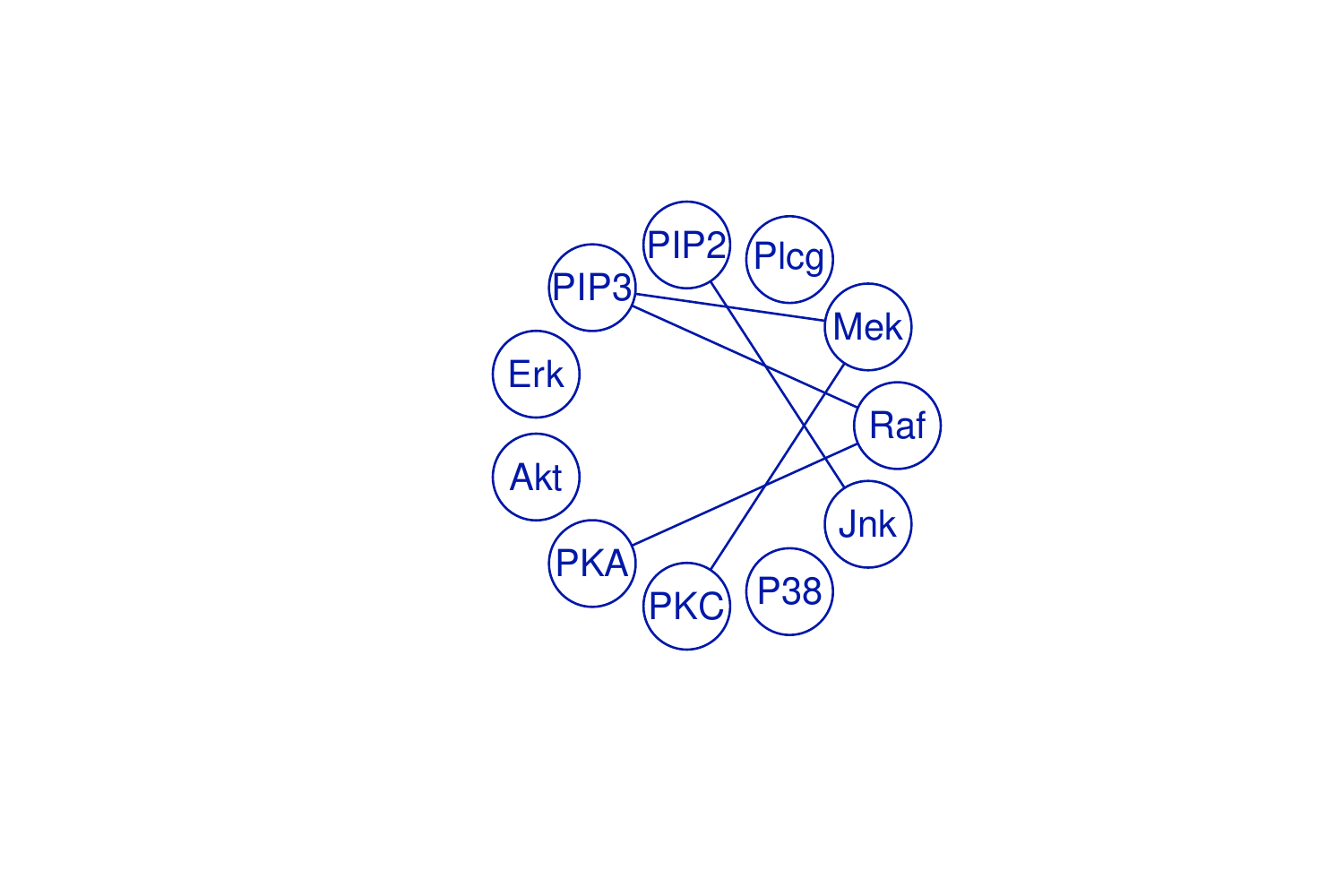}}\hfill
   \subcaptionbox{$\pi^{(1,0)}(G)$ with no extra edges}{\includegraphics[width=0.5\textwidth]{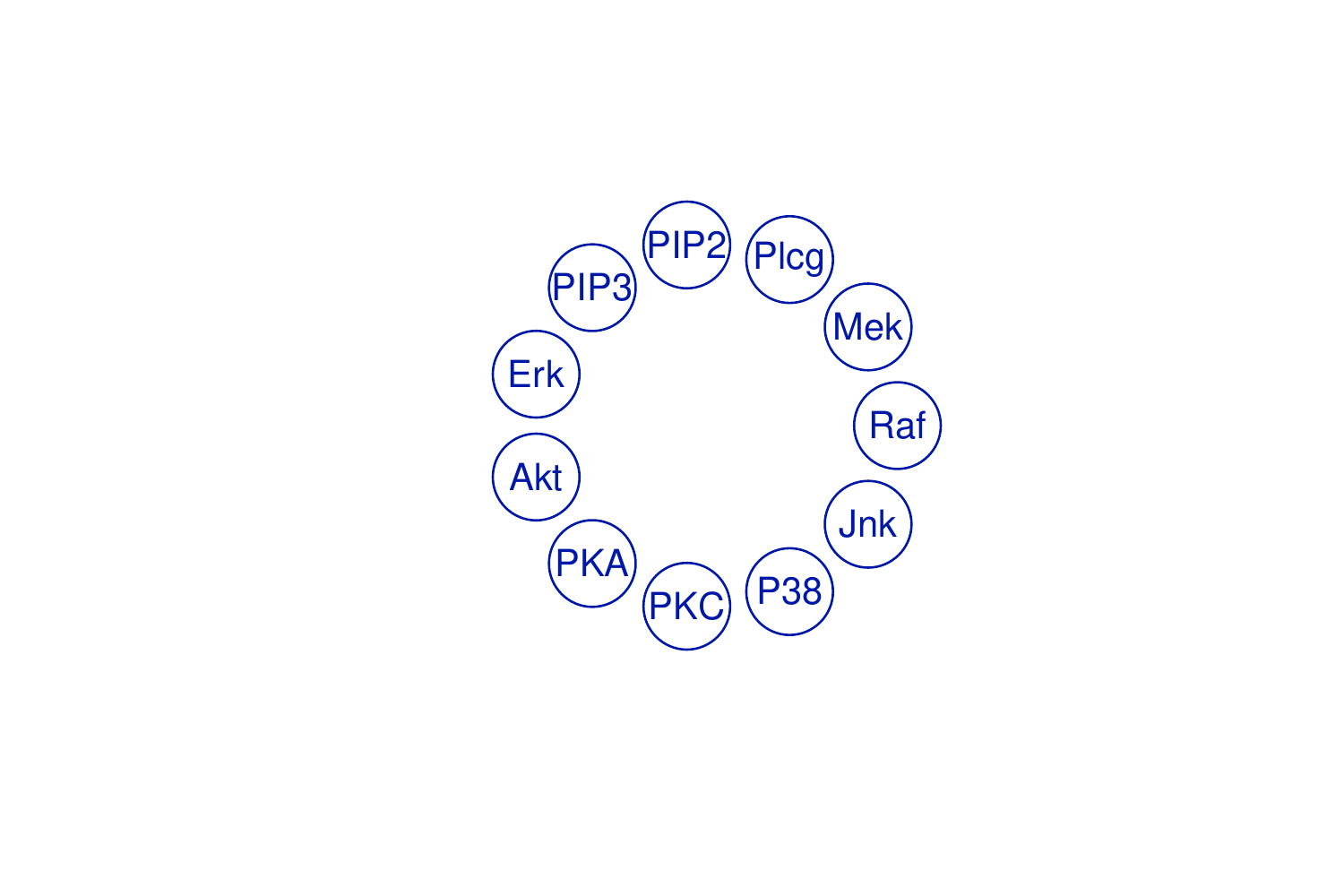}}
    \end{figure}
\begin{figure}[h!]
\ContinuedFloat
\centering
   \subcaptionbox{$\pi^{(1,1/2)}(G)$ with 1 extra edge}{\includegraphics[width=0.5\textwidth]{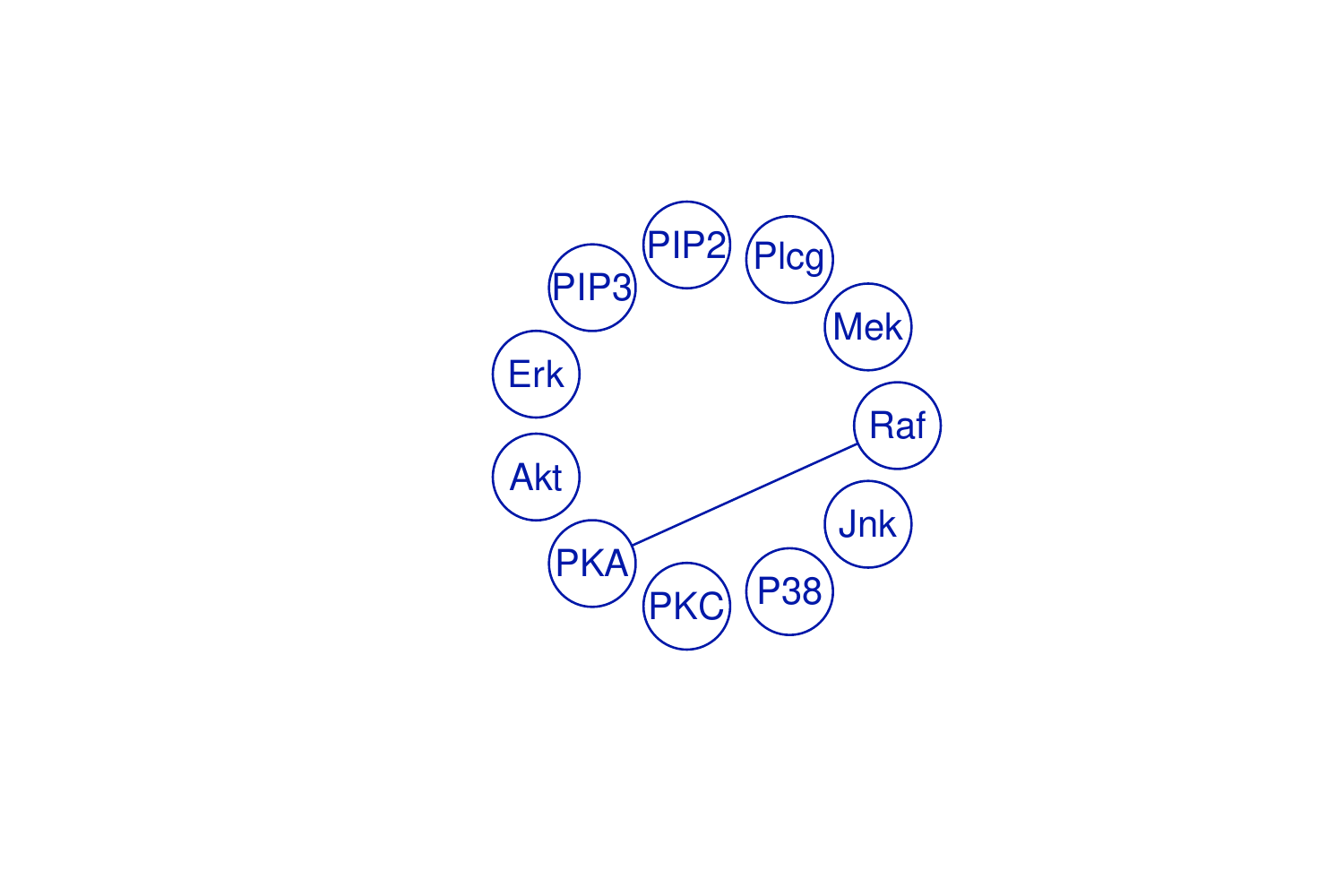}}\hfill
     \subcaptionbox{$\pi^{(0,c)}(G)$ with 2 extra edges} {\includegraphics[width=0.5\textwidth]{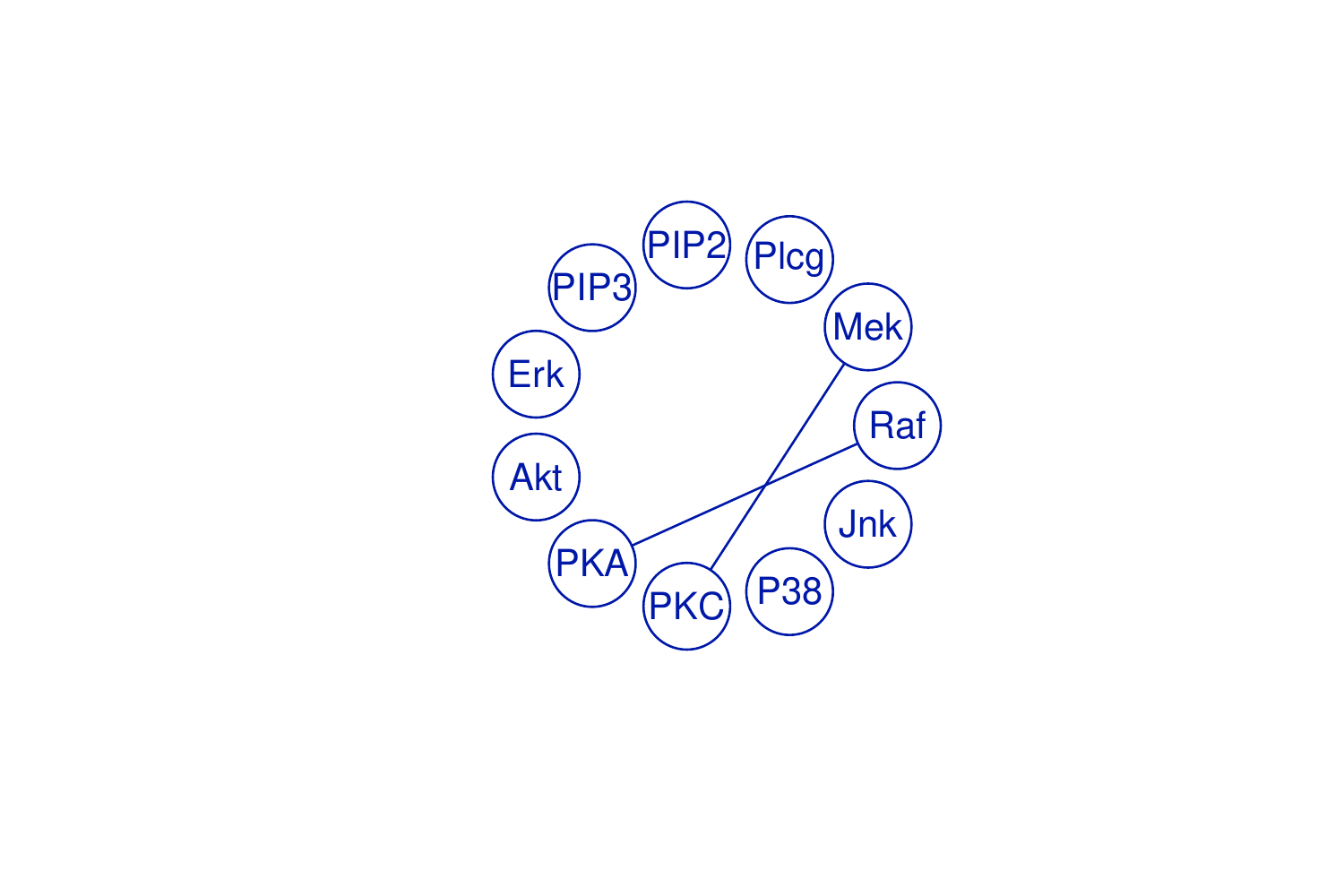}}
     \caption{Estimated graphs for the flow cytometry dataset under each considered prior. Panel (a) shows the edges common to all the priors. Panels (b) to (e) show the additional edges peculiar to each prior.}\label{Fig:FlowCyto}
 \end{figure}
  
\subsubsection{The PTSD Symptoms for Earthquake Survivors in Wenchuan, China Dataset}\label{sc_earthquake}

This dataset \citep{McNally2015} represents the measurement of 17 symptoms associated with PTSD (Post-traumatic stress disorder) reported by 362 survivors of an earthquake from the Wenchuan county in the Sichuan province, China. Each of the participants indicated through a ordinal scale from 1 to 5 how affected they were by every single one of the 17 PTSD symptoms, where 1 indicates not being bothered by the symptom at hand, whereas 5 corresponds to an extreme response to the same symptom. All participants have lost at least one child in the respective earthquake. The data is available with the \texttt{R} package APR \citep{Mair2015}. Amongst those 362 answers, in 18 cases, there was missing information associated with one or several symptoms. These cases were discarded, leaving a final sample of 344 participants, and the data was centred.

\begin{figure}[h!]
  \centering
    \subcaptionbox{The number of common edges is 43}{\includegraphics[width=0.5\textwidth]{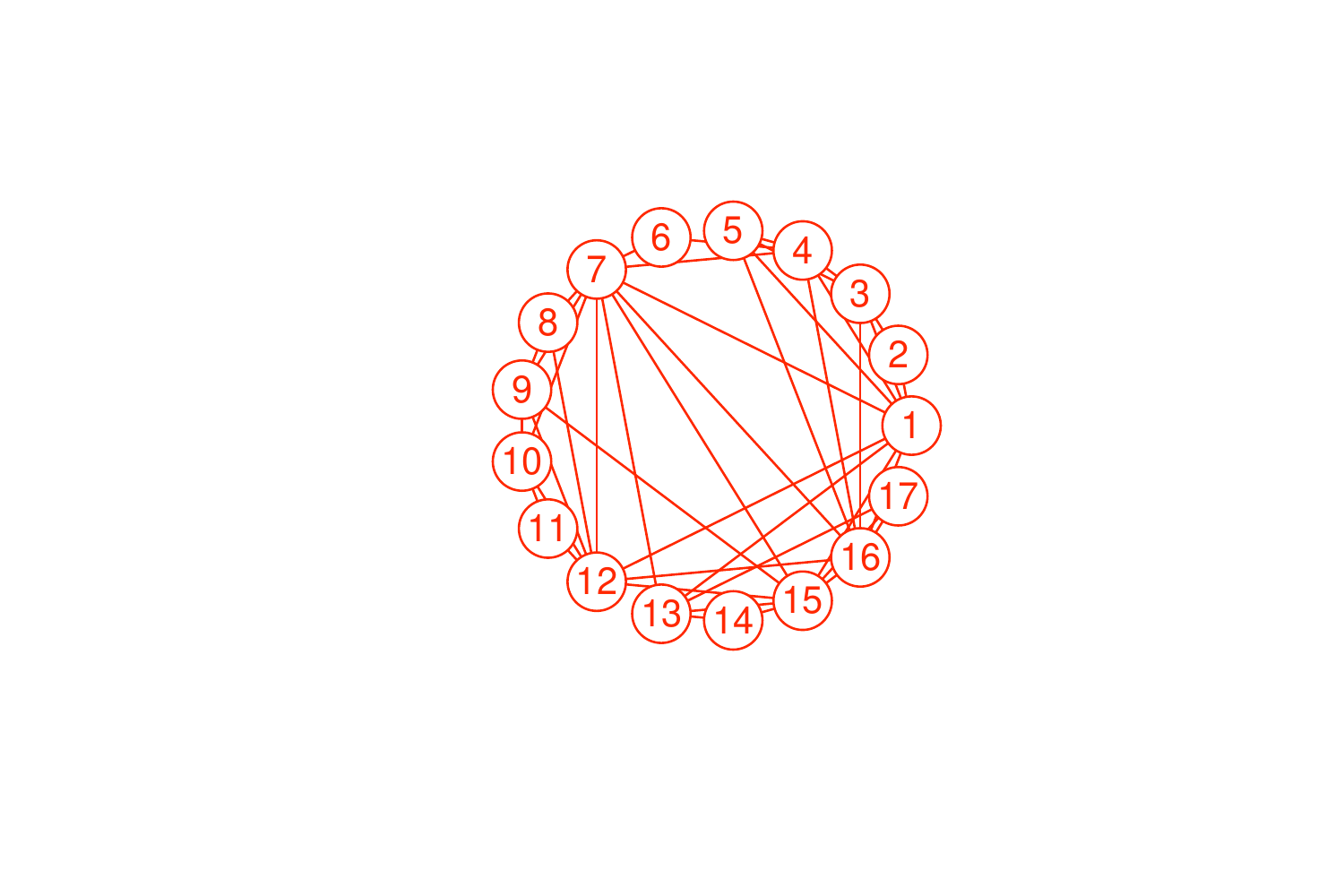}}
  \subcaptionbox{$\pi^{(1,1)}(G)$ with 6 extra edges}{\includegraphics[width=0.5\textwidth]{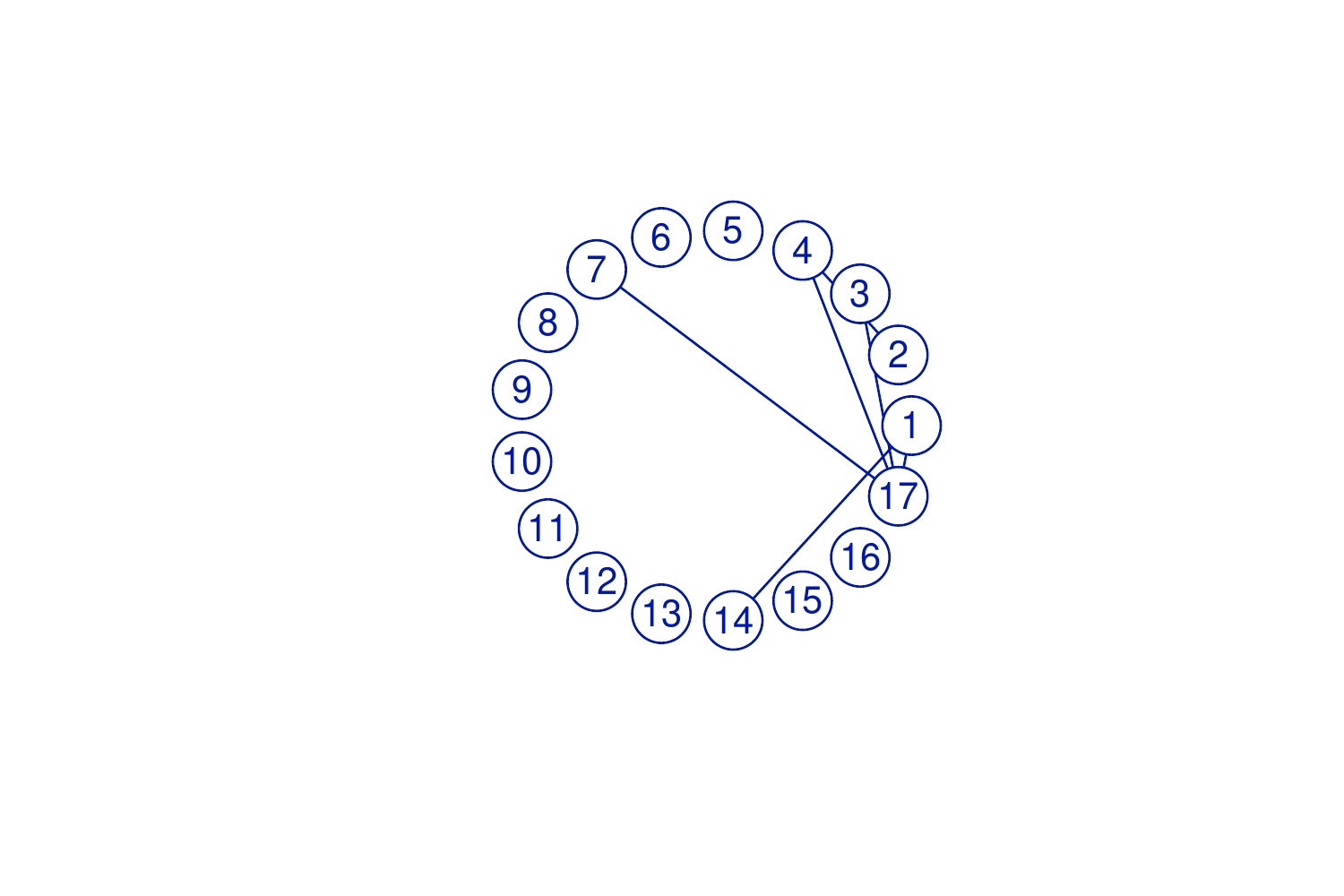}}\hfill
   \subcaptionbox{$\pi^{(1,0)}(G)$ with 5 extra edges}{\includegraphics[width=0.5\textwidth]{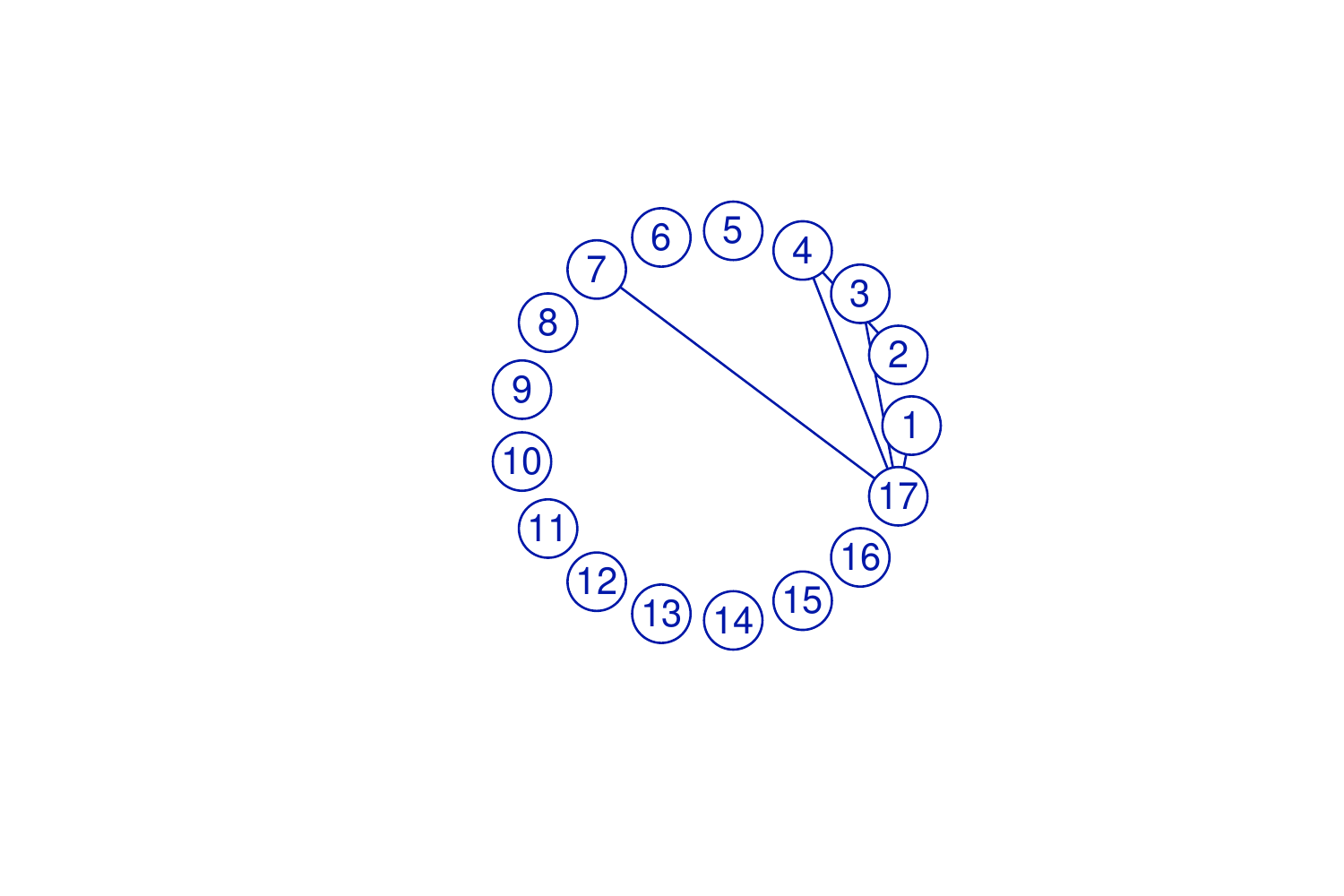}}
   \subcaptionbox{$\pi^{(1,1/2)}(G)$ with 1 extra edge}{\includegraphics[width=0.5\textwidth]{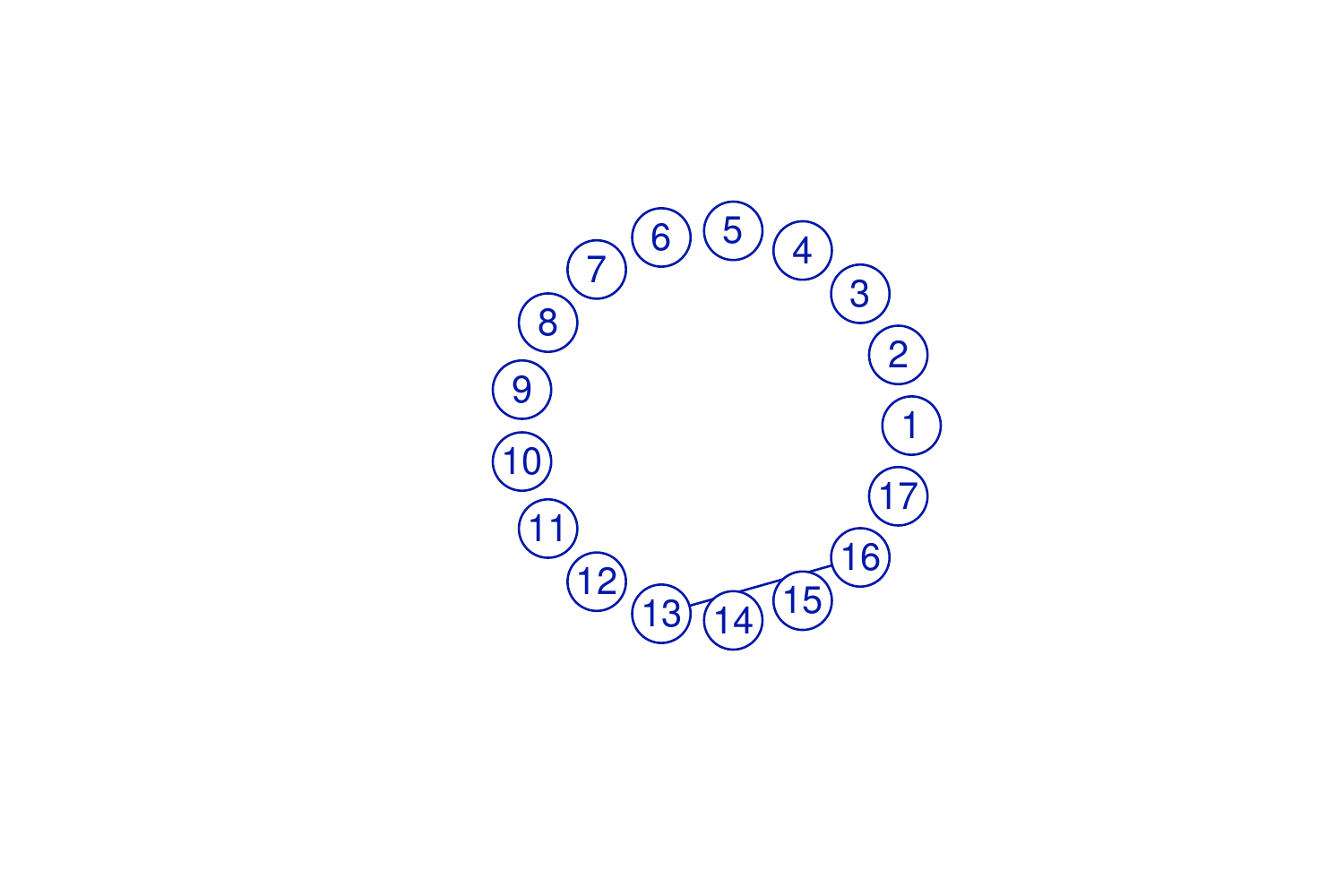}}\hfill
     \subcaptionbox{$\pi^{(0,c)}(G)$ with 6 extra edges} {\includegraphics[width=0.5\textwidth]{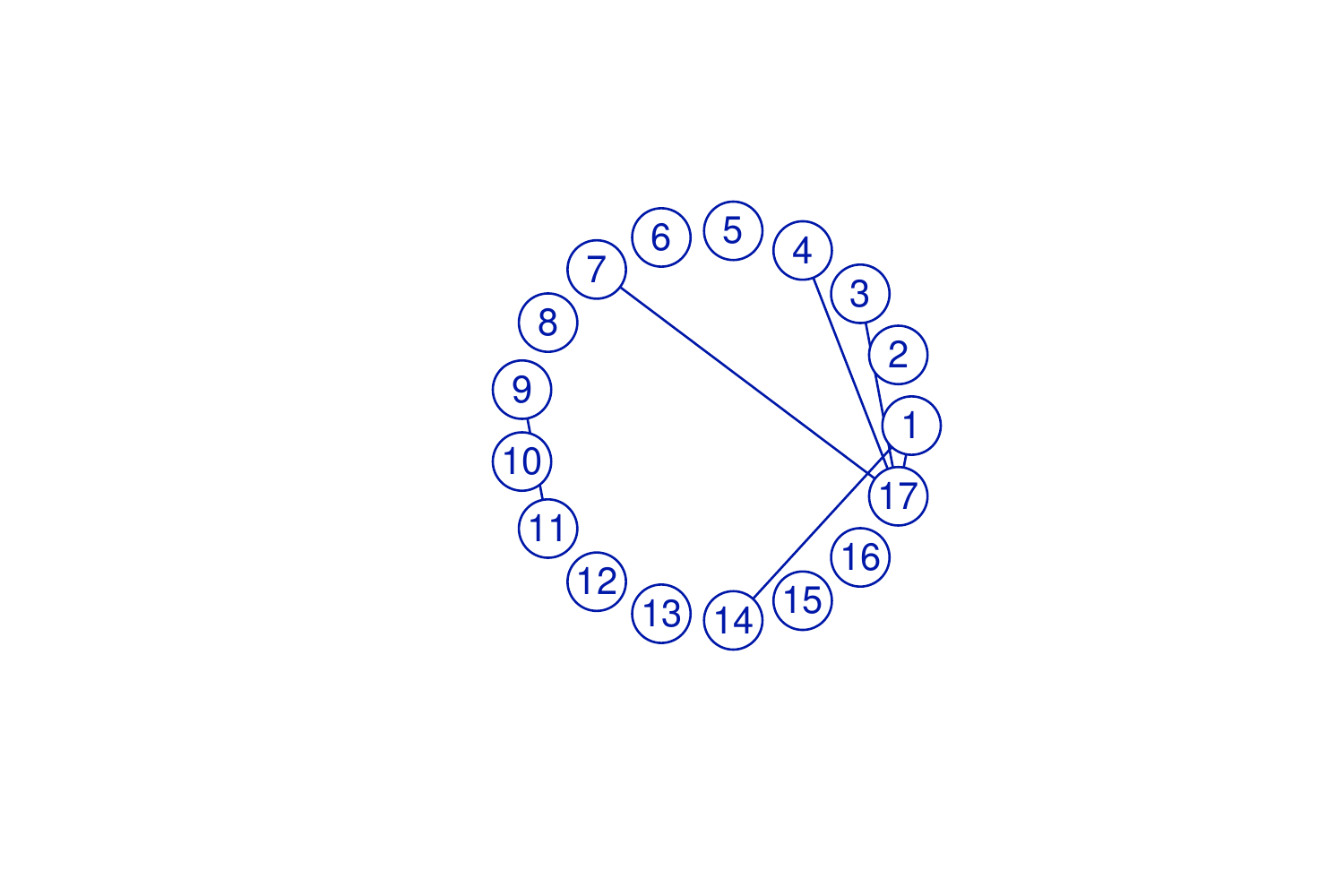}}
     \caption{Estimated graphs for the PTSD dataset under each considered prior. Panel (a) shows the edges common to all the priors. Panels (b) to (e) show the additional edges peculiar to each prior (normality assumption).}
     \label{Wenchuan_FINCS_results}
 \end{figure}

First we have assumed normaility, therefore we have applied the FINCS algorithm. As observed in Figure \ref{Wenchuan_FINCS_results}, the sparser graph is identified under the $\pi^{(1,1/2)}(G)$ prior and it contains 44 edges. With exception of edge $(13,16)$, the remaining 43 edges were also included in the other three priors. Table \ref{Wenchuan_omitted} reports the 8 edges not included in all the four priors.   

\begin{tabularx}{\textwidth}{cccccc}
  \hline
Index  & Edge & $\pi^{(1,1)}(G)$  & $\pi^{(1,0)}(G)$ & $\pi^{(1,1/2)}(G)$ & $\pi^{(0,c)}(G)$ \\ 
  \hline 
  1 & (1,14) & 0.608 & 0.492 & 0.413 & 0.763 \\ 
  2 & (1,17) & 1.000 & 1.000 & 0.456 & 1.000 \\ 
  3 & (2,4) & 0.513 & 0.512 & 0.385 & 0.463  \\ 
  4 & (3,17) & 0.528 & 0.531 & 0.246 & 0.634 \\ 
  5 & (4,17) & 0.994 & 0.969 & 0.442 & 0.998 \\ 
  6 & (7,17) & 0.908 & 0.895 & 0.414 & 0.999 \\ 
  7 & (9,11) & 0.495 & 0.405 & 0.431 & 0.663 \\ 
  8 & (13,16) & 0.027 & 0.019 & 0.562 & 0.045 \\ 
\caption{Edges with a posterior inclusion probability larger than 0.5 for one to three of the four considered priors.}
\label{Wenchuan_omitted}
\end{tabularx}


\begin{figure}[h!]
  \centering
  \subcaptionbox{The number of common edges is 5}{\includegraphics[width=0.5\textwidth]{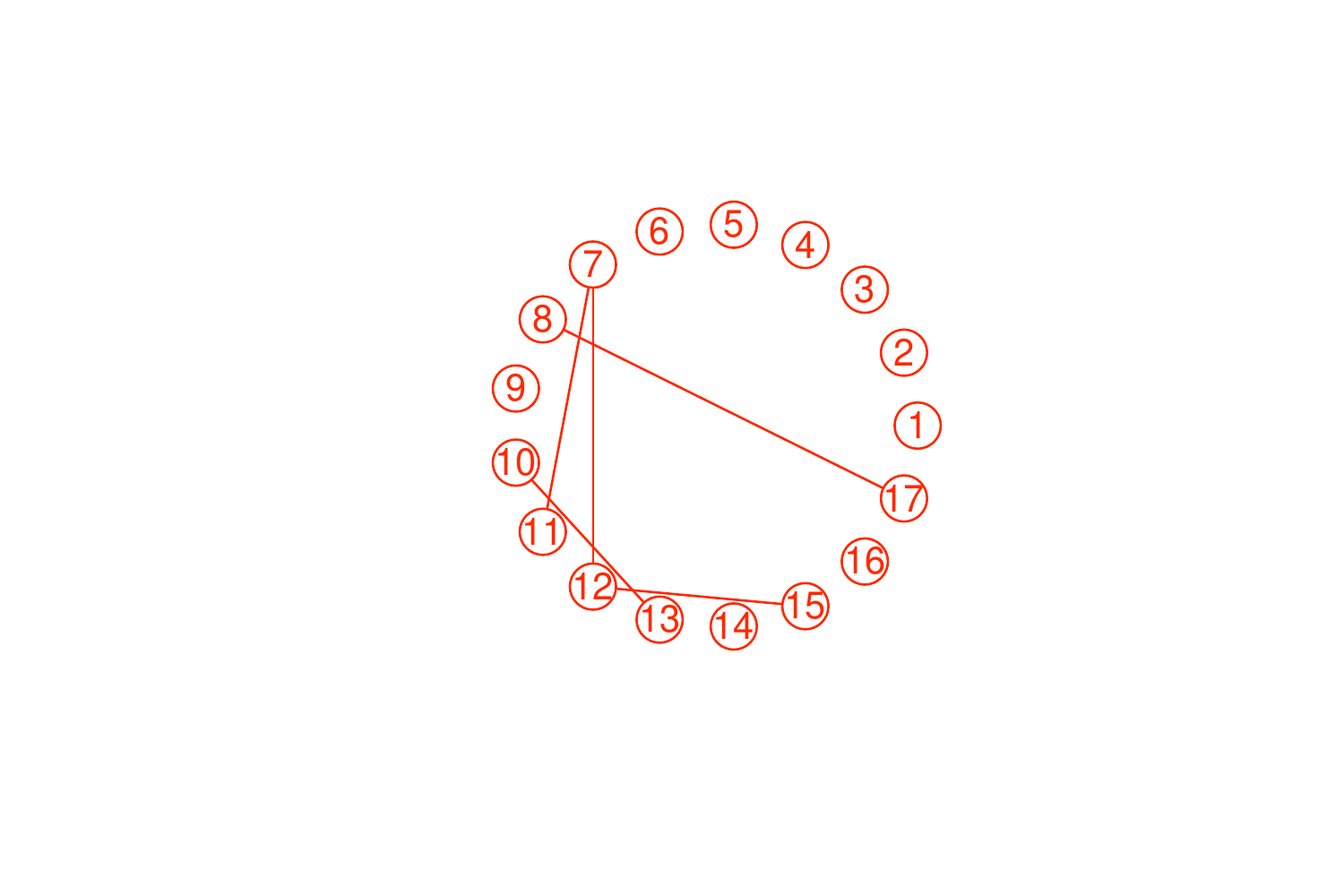}}\hfill
  \subcaptionbox{BD with 54 extra edges}{\includegraphics[width=0.5\textwidth]{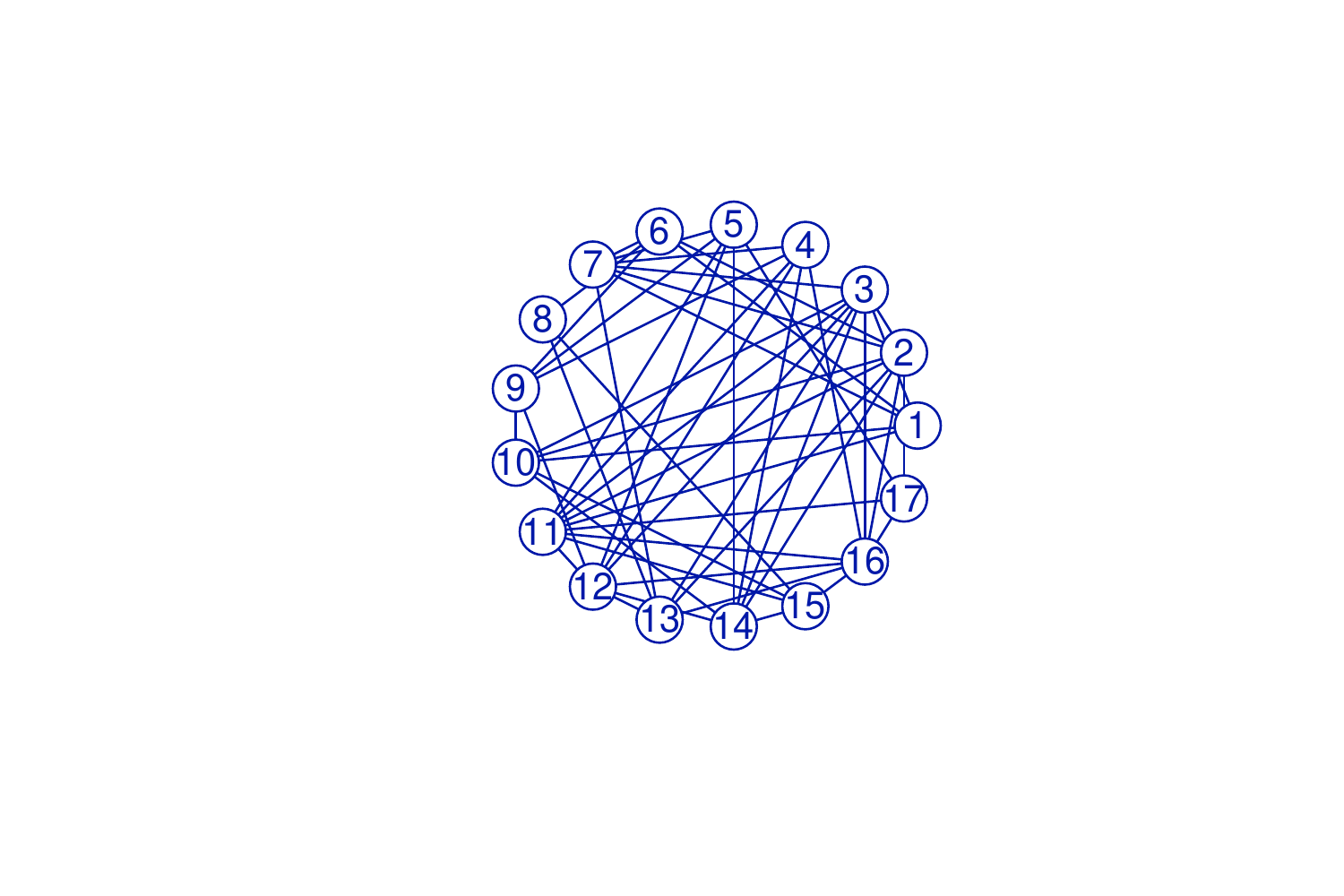}\label{Wenchuan_BD}}
   \subcaptionbox{$\pi^{(1,0)}(G)$ with 53 extra edges}{\includegraphics[width=0.5\textwidth]{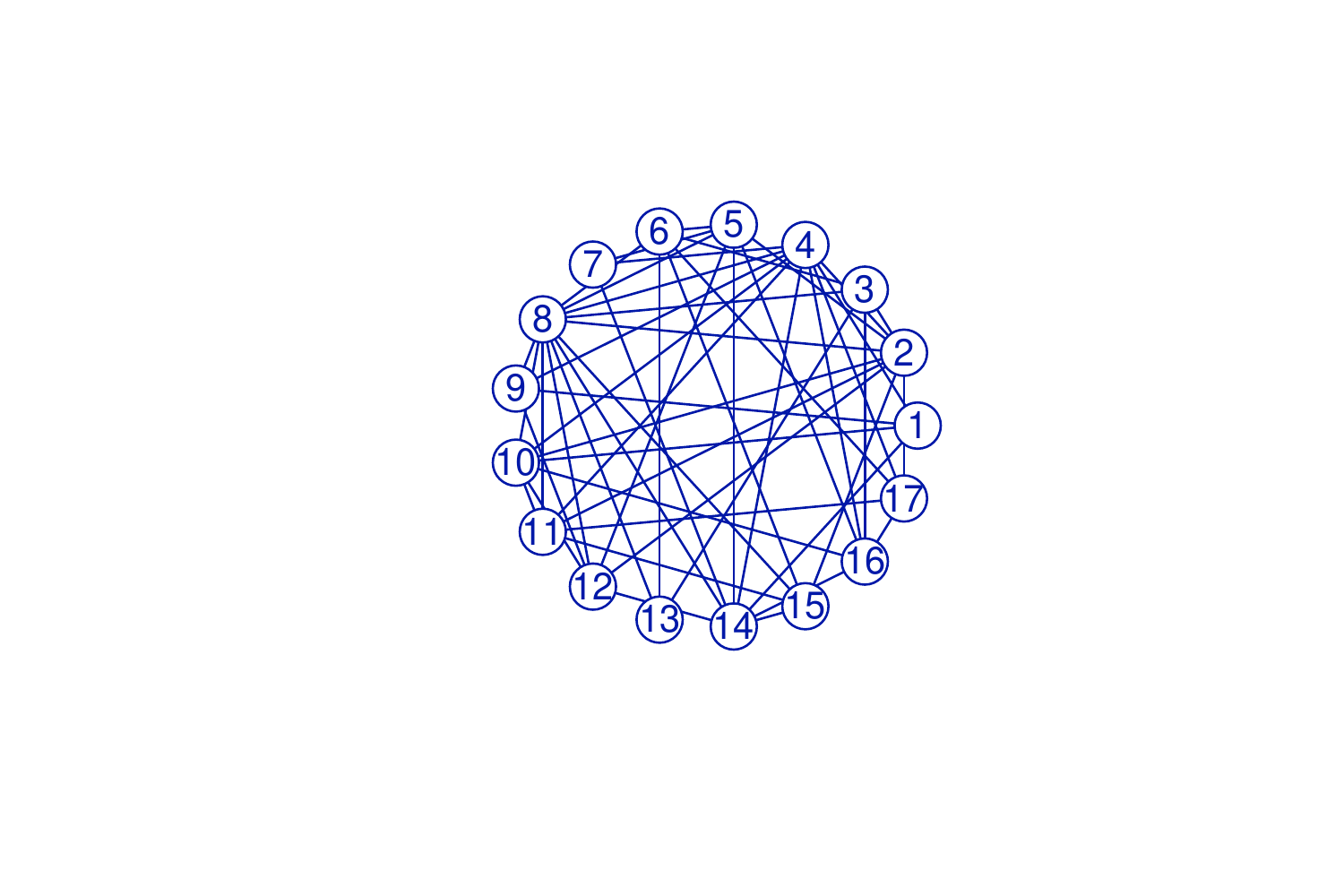}}\hfill
   \subcaptionbox{$\pi^{(1,1/2)}(G)$ with 63 extra edges}{\includegraphics[width=0.5\textwidth]{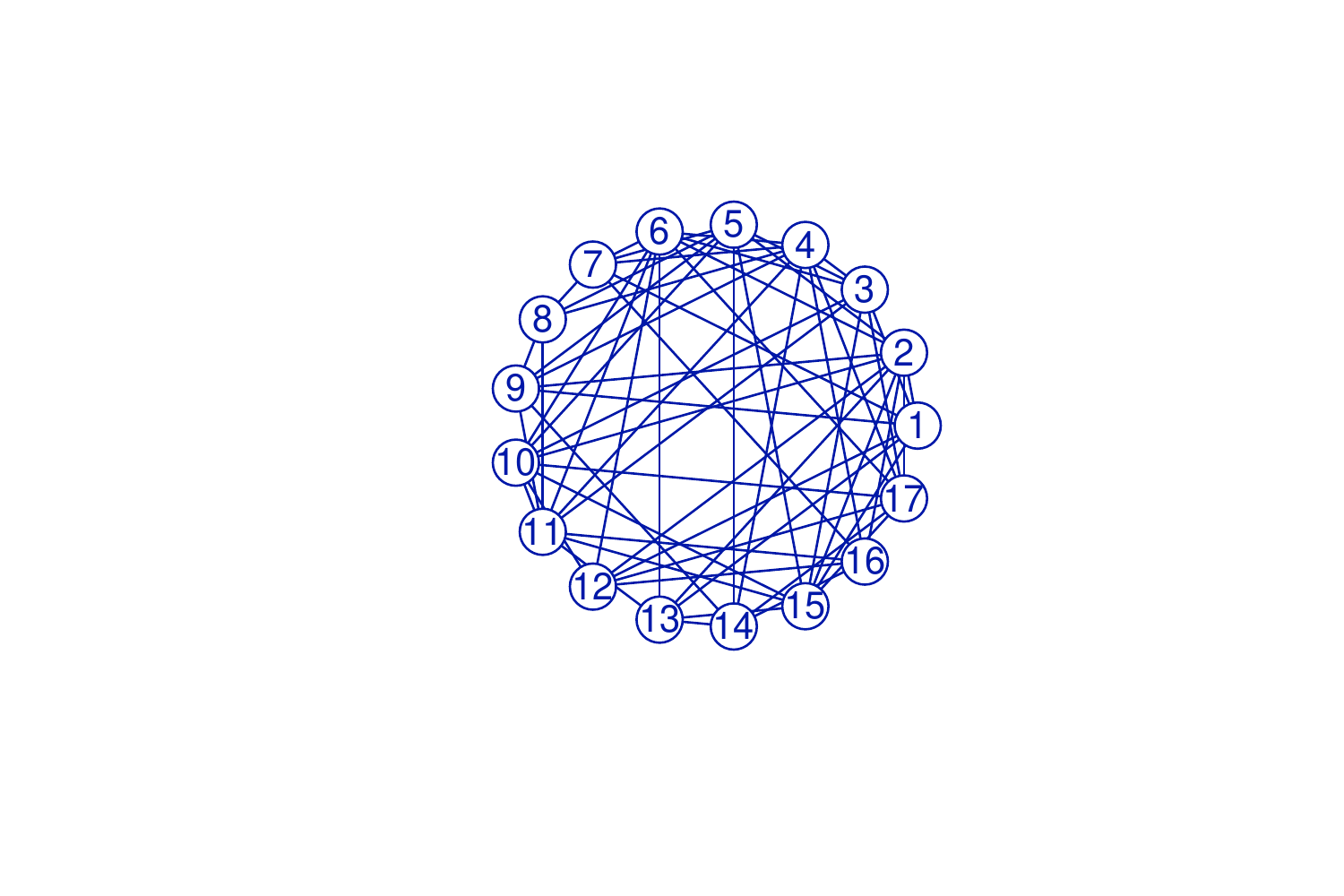}}
     \subcaptionbox{$\pi^{(1,1)}(G)$ with 54 extra edges} {\includegraphics[width=0.5\textwidth]{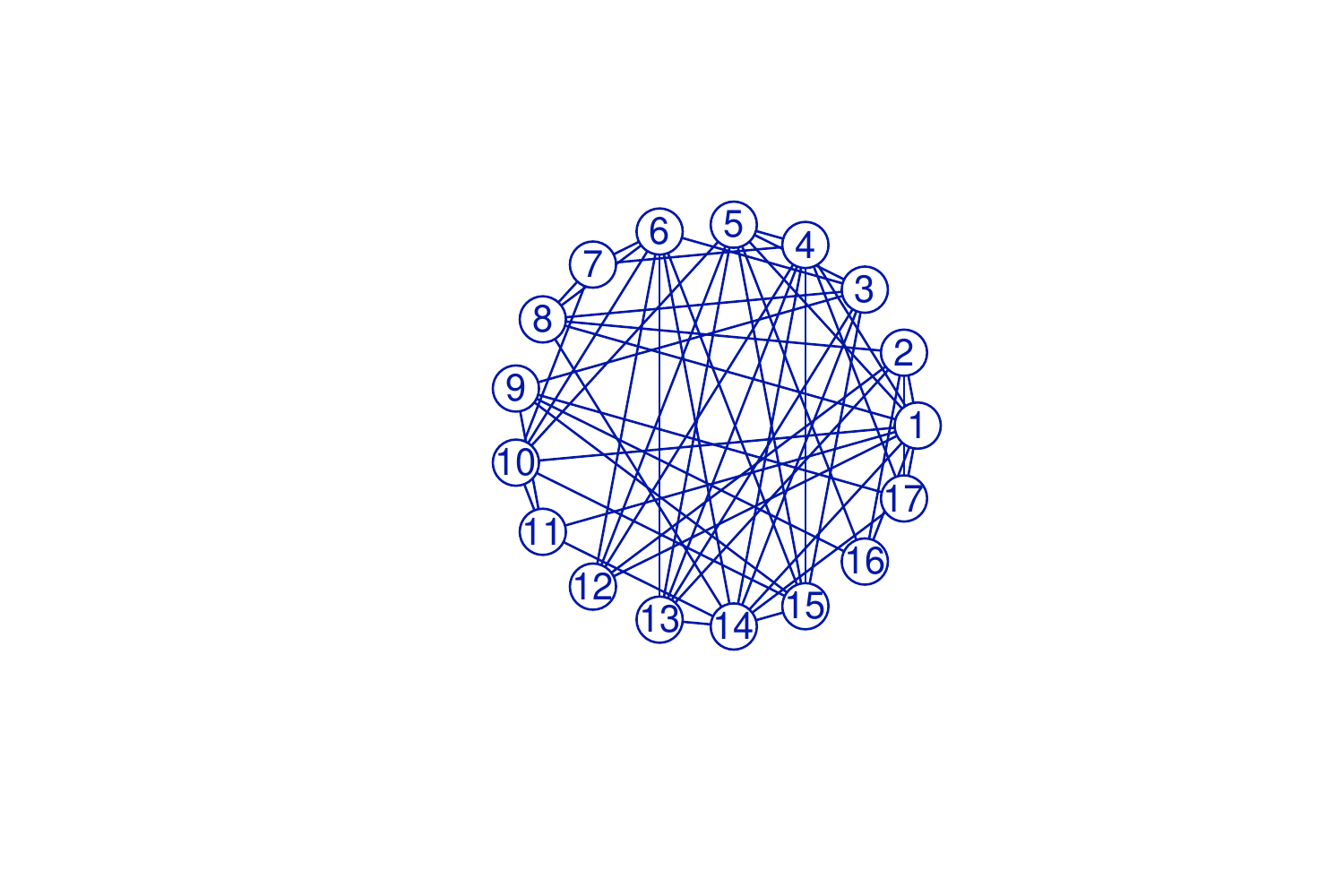}}\hfill
   \subcaptionbox{$\pi^{(0,c)}(G)$ with 66 extra edges}{\includegraphics[width=0.5\textwidth]{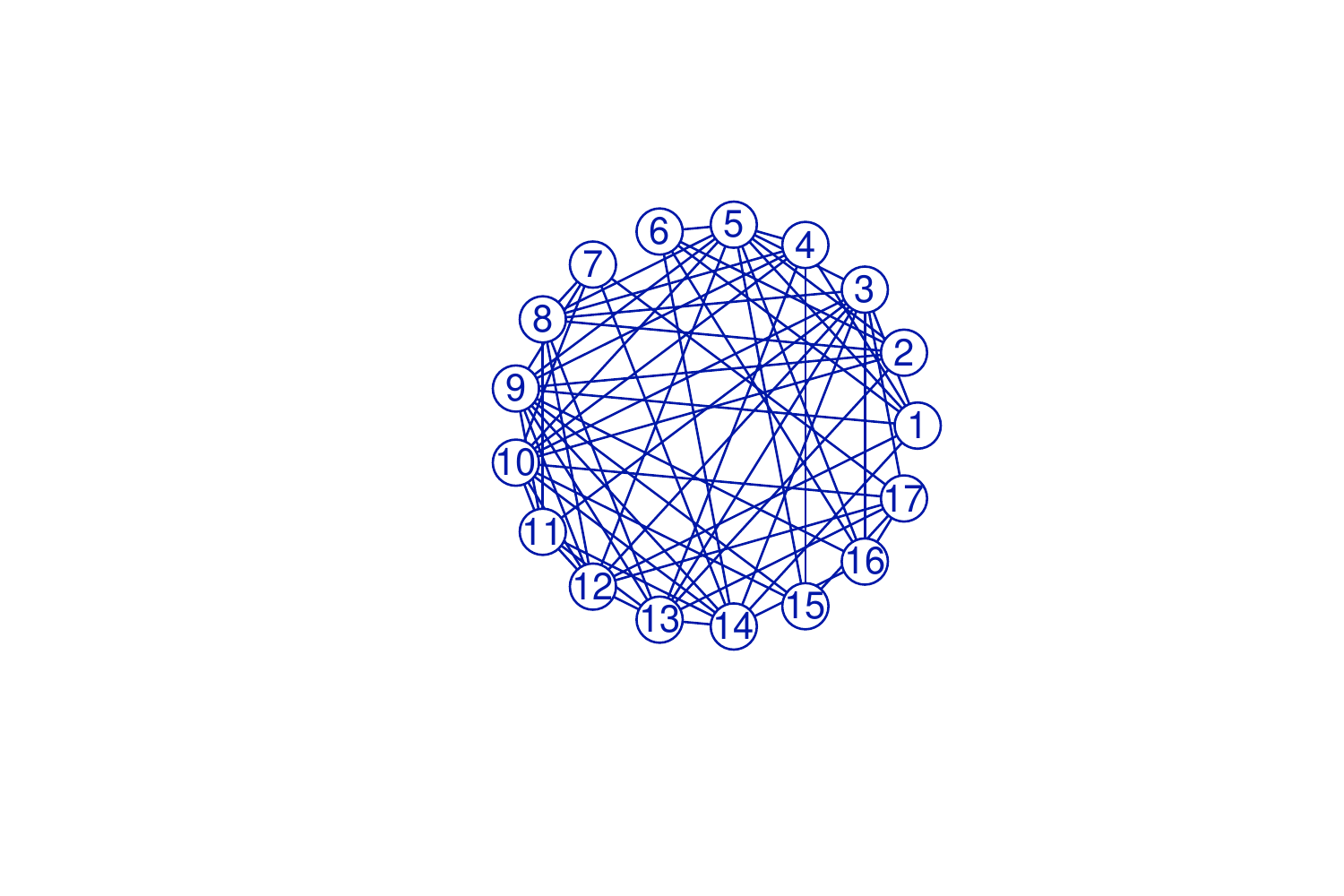}}
  \caption{Estimated graphs for the PTSD under each considered prior. Panel (a) shows the edges common to all the priors. Panels (b) to (e) show the additional edges peculiar to each prior (copula approach).}
  \label{Wenchuan_MAP_graph_sizes}
\end{figure}

We have then relaxed the normality assumption by using the copula approach implemented in the  BDgraph package, \citep{MohammadiCopula2017}. We have compared the maximum posterior graphs which were found under the $\pi^{(1,1)}(G)$, $\pi^{(1,0)}(G)$, $\pi^{(0,c)}(G)$ and a specific case of the mixture prior where the parameters were set to $h=1$ and $c=0.5$  when $2000000$ iterations were used with the first $1000000$ discarded. We have also included the Bernoulli prior (denoted with BD) where the probability of edge inclusion was set to 0.004 so that a priori it had the same mean as the $\pi^{(1,0)}(G)$ prior, namely 0.582. The maximum posterior probability graph found under the Bernoulli prior has a size of 59 and can be seen in Figure \ref{Wenchuan_MAP_graph_sizes}. Furthermore, as we can observe in Figure \ref{Wenchuan_MAP_graph_sizes}, under the $\pi^{(1,1)}(G)$ prior, the posterior graph had a size of 59, whilst for the $\pi^{(1,0)}(G)$ prior, the posterior had a size of 58. When we use the mixture prior, the size of the maximum posterior probability graph is 68. The most dense graph is obtained under the $\pi^{(0,c)}(G)$ which had a size of 71. The number of common edges between the BD and various graphs is as following: 20 edges with the $\pi^{(1,1)}(G)$, 29 edges with the $\pi^{(1,0)}(G)$, 27 with the $\pi^{(1,1/2)}(G)$ and $26$ with the $\pi^{(0,c)}(G)$. Clearly, under this particular running of the bdmcmc algorithm, the sparser graph is found under the $\pi^{(1,0)}(G)$ prior, but the other maximum posterior probability graphs corresponding to the BD and $\pi^{(1,1)}(G)$ priors are close to its size. The only exception is given by the uniform prior which leads to a denser graph.

\subsubsection{The Breast Cancer Dataset}\label{sc_breastcancer}
\citet{Hess2006} have collected gene expression data for 133 patients which had breast cancer. This dataset was also analysed by \citet{Ambroise2009} and made available through the \texttt{R} package SIMONE (Statistical Inference for MOdular NEtworks) \citep{SIMONE}. There are 26 genes considered in the study. The dataset is split in two groups, one pertaining to the pathological complete response (pCR) to the chemotherapy treatment started after surgery, whereas the other corresponds to the disease still being present in the patients (not-pCR). We have focused our analysis on the pCR group only, and centred the data. 

\begin{figure}[h!]
  \centering
    \subcaptionbox{The number of common edges is 17}{\includegraphics[width=0.5\textwidth]{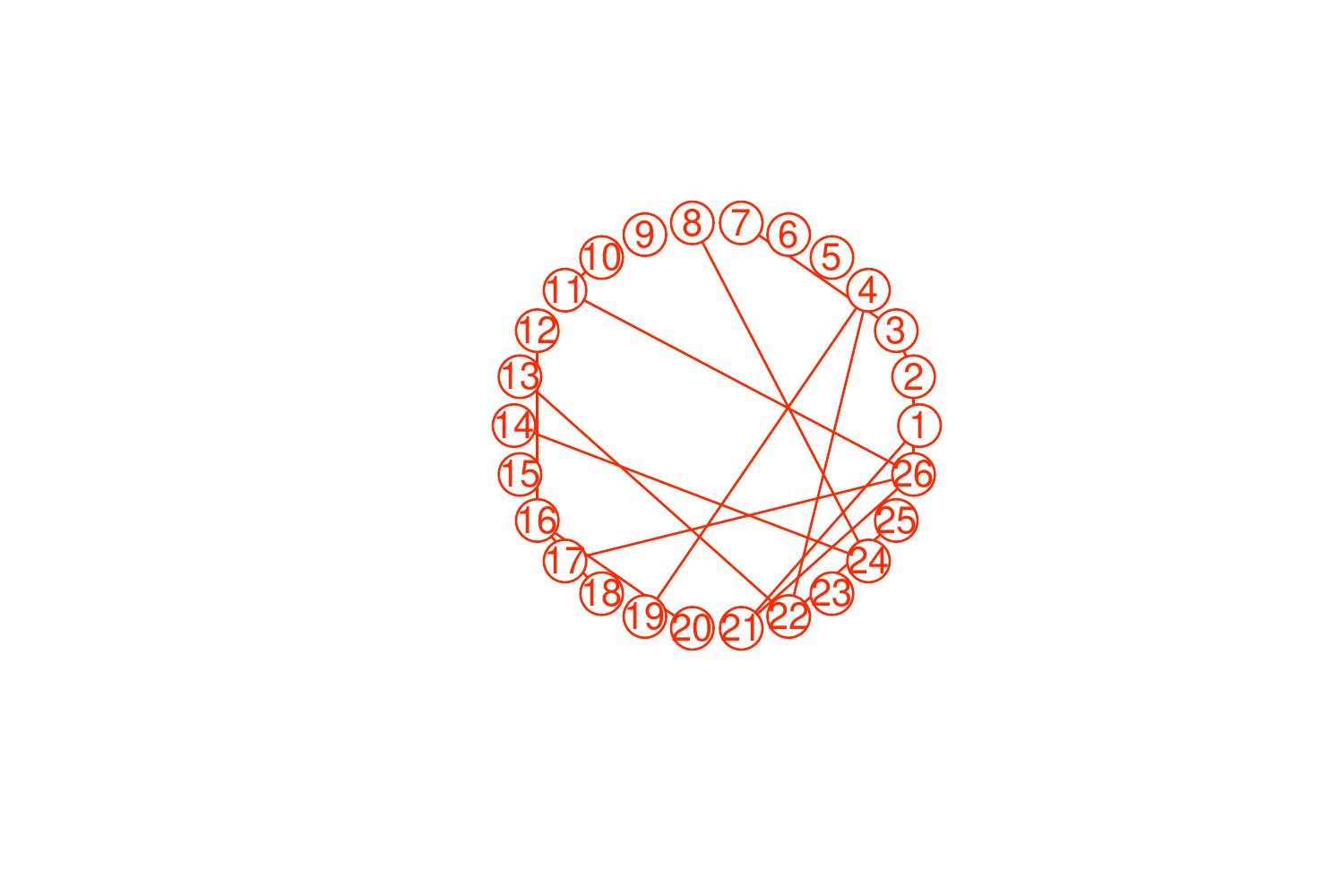}}
  \subcaptionbox{$\pi^{(1,1)}(G)$ with 4 extra edges}{\includegraphics[width=0.5\textwidth]{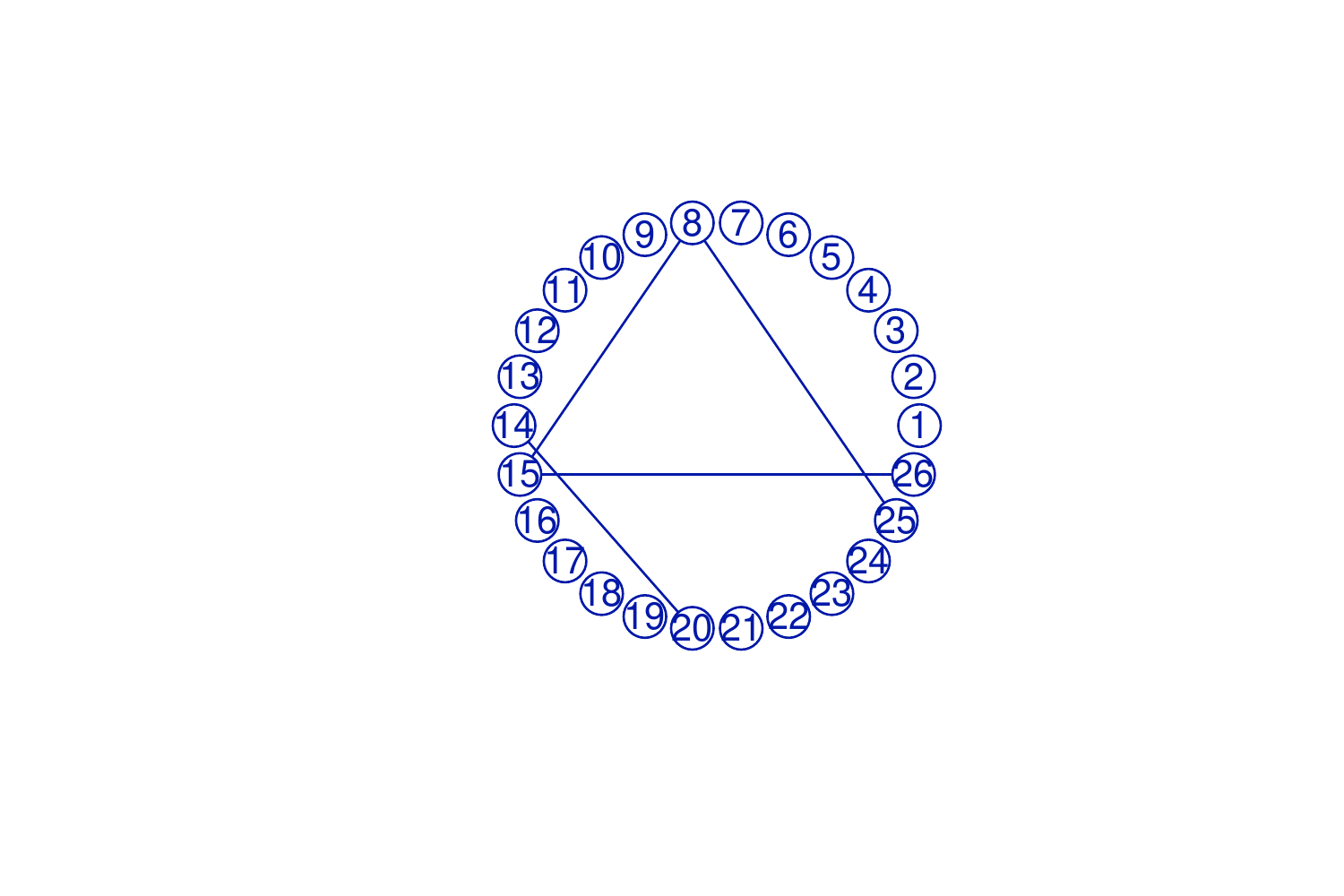}}\hfill
   \subcaptionbox{$\pi^{(1,0)}(G)$ with 11 extra edges}{\includegraphics[width=0.5\textwidth]{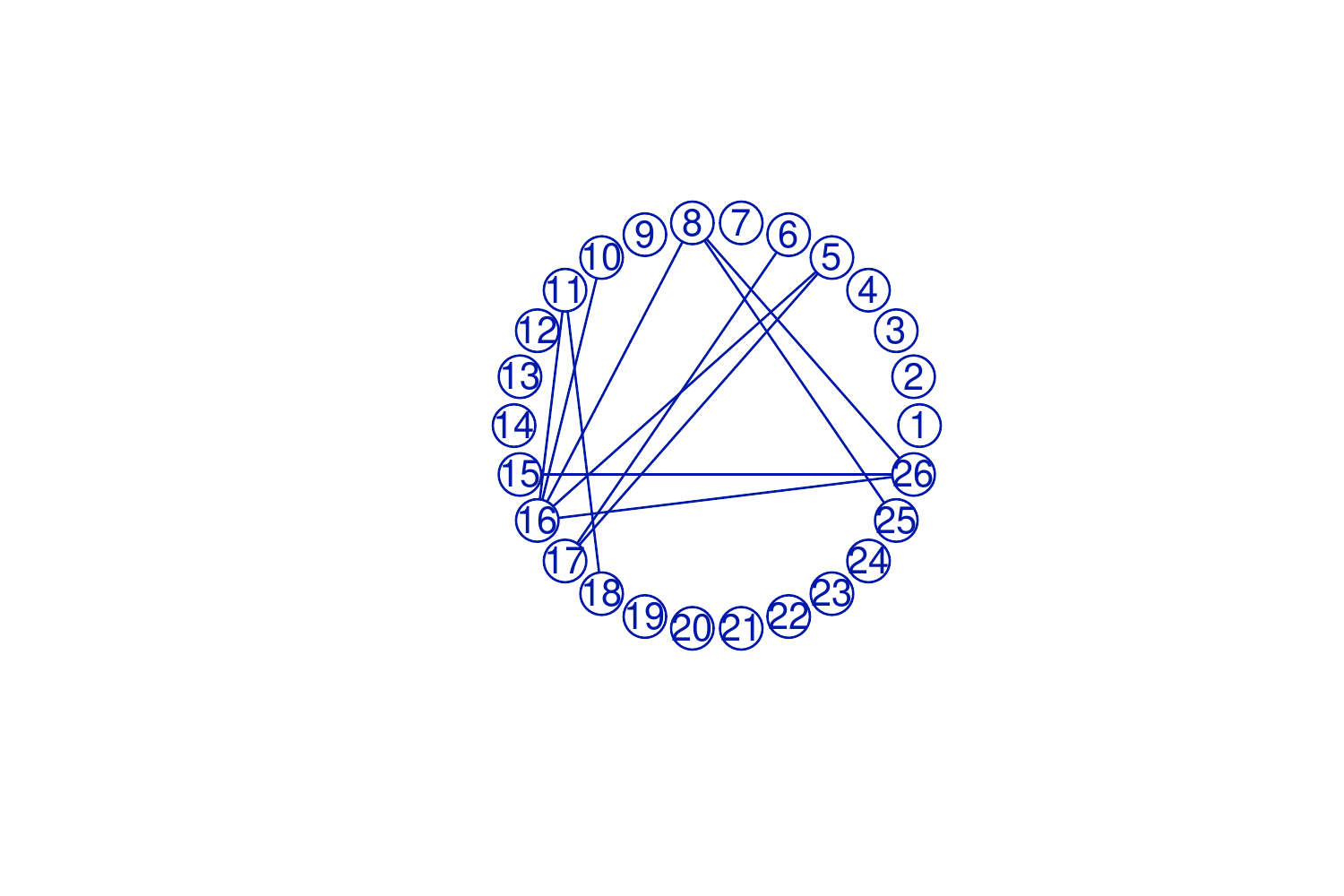}}
   \subcaptionbox{$\pi^{(1,1/2)}(G)$ with 6 extra edges}{\includegraphics[width=0.5\textwidth]{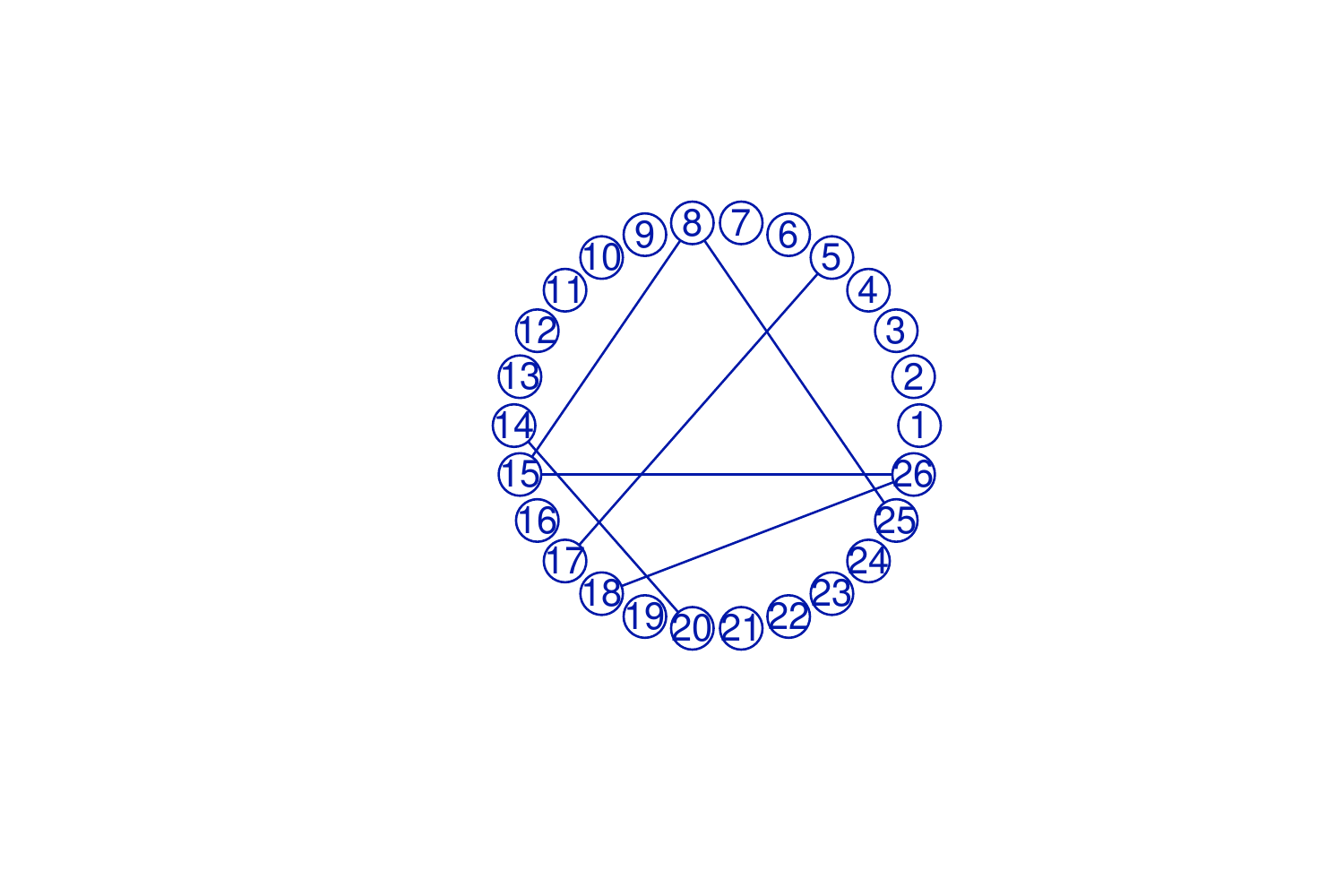}}\hfill
     \subcaptionbox{$\pi^{(0,c)}(G)$ with 25 extra edges} {\includegraphics[width=0.5\textwidth]{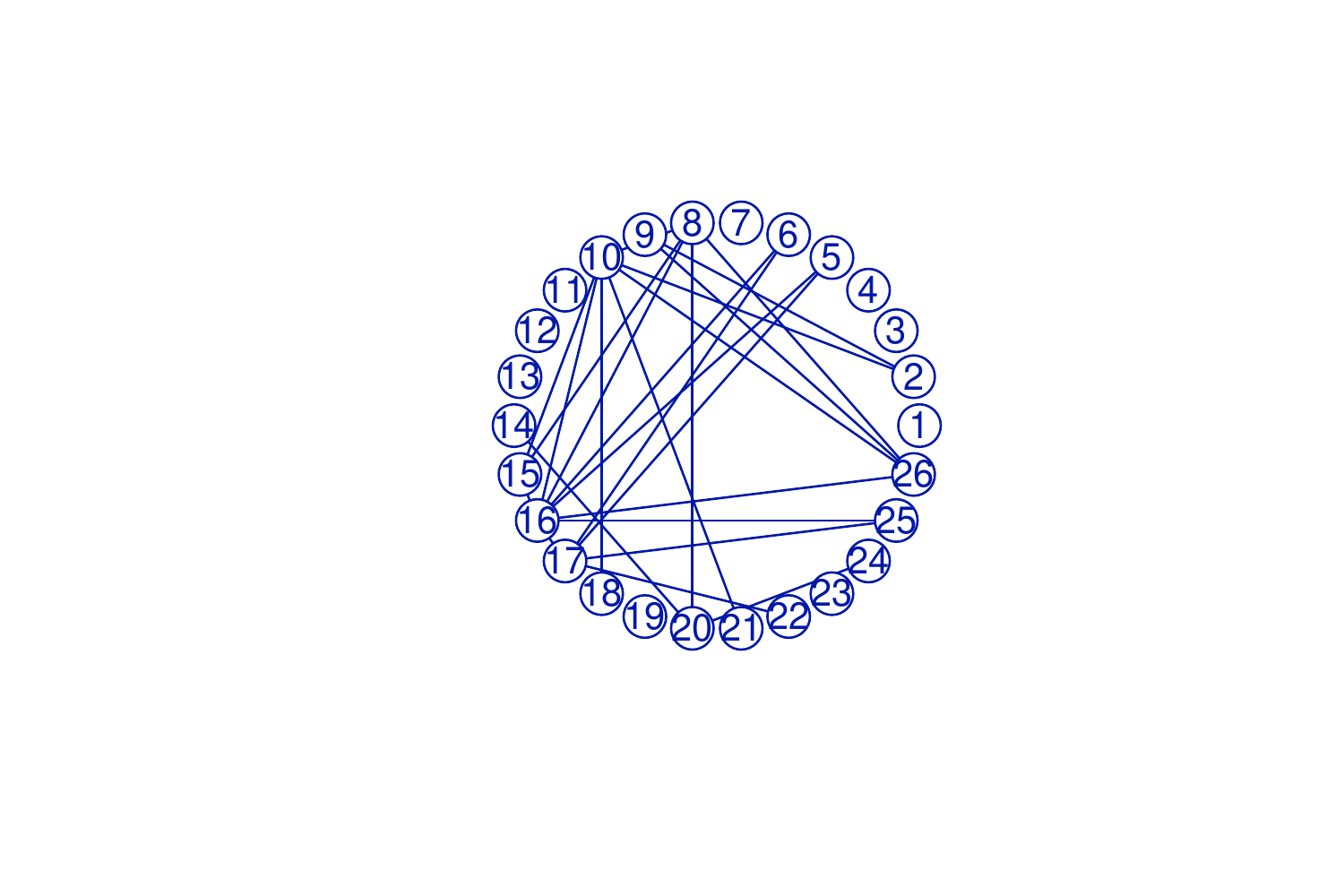}}
     \caption{Estimated graphs for the pCR group in the Breast Cancer dataset under each considered prior. Panel (a) shows the edges common to all the priors. Panels (b) to (e) show the additional edges peculiar to each prior.}
    \label{fig:cancerdata}
 \end{figure}

The estimated graphs are reported in Figure \ref{fig:cancerdata}, where we have shown the results under each prior, that is $\pi^{(1,1)}(G)$, $\pi^{(1,0)}(G)$, $\pi^{(1,1/2)}(G)$ and $\pi^{(0,c)}(G)$, as well as the graph of the edges common to all the priors. Comparing the performance of the priors, we note that the $\pi^{(1,1)}(G)$ prior and the $\pi^{(1,1/2)}(G)$ prior give relatively sparse graphs, 21 and 23 edges respectively. The $\pi^{(1,0)}(G)$ prior yields a slightly larger graph (28 edges) while the $\pi^{(0,c)}(G)$ is the prior resulting in the most complex graph (42 edges). To further ease the comparison of the priors, we have reported in Table \ref{Hess_pCR_omitted} the posterior inclusion probabilities of the edges not included in all the graphs. We can see a confirmation of the above results as the edges non included in all the prior tend, with few exception, to have a posterior inclusion probability larger than 0.5 under the $\pi^{(1,0)}(G)$ prior and/or the $\pi^{(0,c)}(G)$ prior.

\section{Conclusion and discussion}\label{sc_conclusions}
In the present work, we have illustrated a novel prior for the space of graphs in the context of Graphical Gaussian Models. The prior is derived using a loss with two components: one relative to the informational content of the graph and one related to its complexity. As displayed in equation \ref{RefinedOurPrior}, the prior has  the form:
 \begin{equation}
\pi^{(h,c)}(G) \propto \exp\left\lbrace-h\left[(1-c)|G|+c\log\binom{m}{|G|}\right]\right\rbrace.
\end{equation} 
We would like to provide some general remarks on the setting of the parameters $h$ and $c$ of the proposed prior. There are several ways to approach the issue:

\begin{itemize}
\item One could set $h$ and $c$ to reflect prior information. See the example where we compare the proposed prior to the Bernoulli prior in Section 4.1.
\item An alternative choice is to set $c=0$ so that the prior will reduce to the global loss component only. Here, the parameter $h$ can be either set according to some prior information or in a default manner (see Villa and Lee, 2019).
\item The third choice is to set $c=1$ and $h=1$ and obtain Carvalho and Scott's prior. This would be the choice if one is interested in multiplicity correction.
\item Finally, one could fix $h=1$ and then set $c$ so to have a desired balance between the global and the local loss due to complexity. We have suggested that a default choice is for $c=0.5$. In this scenario as well, given that the prior will depend on the total number of edges, there is correction for multiplicity.
\end{itemize}

We applied the above prior for different combinations of $h$ and $c$. The results were obtained by implementing the FINCS algorithm and comparison were made with to alternative weakly informative priors: the uniform prior and the prior prosed in \citet{Carvalho}, both of which can be seen as a particular case of the proposed prior.  

\medskip

Simulation studies, performed under a non-informative assumption, show that the best configuration of the proposed prior is when equal weight is given to absolute and relative complexity. In fact, the results are similar to the $\pi^{(1,1)}(G)$ prior. Simulation studies in scenarios of prior information have provided evidence in favour of the proposed prior, in particular when compared to the Bernoulli prior used by \cite{Moham}. Here we show that the dependence of the prior from two parameters allows to better include initial information when is not limited to one piece only.  One aim of the simulations was to compare the priors in terms of convergence rate and computational time. We have therefore considered cases of 100,000 and 5 millions iterations for graphs of different sizes, noticing that the  $\pi^{(1,1)}(G)$ and the $\pi^{(1,1/2)}(G)$ priors result in faster convergence with respect to the other two priors. Less differences have been noted in the computational times with the exception of the slower performance of the uniform prior. 
 
\medskip

Finally, we have illustrated the prior for three real datasets of different dimensionality and size. The proposed prior, in terms of sparsity, yields results in line with the $\pi^{(1,1)}(G)$ prior, with the clear better performance for the first dataset (Flow Cytometry dataset).  In fact, for this case, we note the posterior log-score for $\pi^{(1,0)}(G)$ is -502974.61 while for the $\pi^{(1,1)}(G)$ prior is -502970.35; note that for the above priors the median probability graph coincides to the highest posterior probability graph. We havo also noted that, when the dimensionality to moderate/high, as it is the case of PTSD and Breast Cancer datasets, the log-score of the  $\pi^{(1,1)}(G)$ is slightly better than the log-scores of $\pi^{(1,0)}(G)$ and $\pi^{(1,1/2)}(G)$; with the latter slightly better than the former. In all cases, the uniform prior $\pi^{(0,c)}(G)$ underperforms when compared to the other priors.  

\medskip

\begin{tabularx}{\textwidth}{cccccc}
  \hline
Index  & Edge & $\pi^{(1,1)}(G)$  & $\pi^{(1,0)}(G)$ & $\pi^{(1,1/2)}(G)$ & $\pi^{(0,c)}(G)$ \\ 
  \hline 
  1 & (2,9)  & 0.001 & 0.008 & 0.004 & 0.538 \\ 
  2 & (2,10) & 0.001 & 0.001 & 0.001 & 0.985 \\ 
  3 & (5,16) & 0.190 & 0.522 & 0.333 & 0.994 \\
  4 & (5,17) & 0.309 & 0.750 & 0.510 & 0.996 \\  
  5 & (6,16) & 0.011 & 0.016 & 0.011 & 0.759 \\ 
  6 & (6,17) & 0.111 & 0.775 & 0.464 & 0.983 \\ 
  7 & (8,10) & 0.000 & 0.000 & 0.000 & 1.000 \\ 
  8 & (8,15) & 0.621 & 0.028 & 0.561 & 1.000 \\
  9 & (8,16) & 0.001 & 0.995 & 0.017 & 1.000 \\ 
  10 & (8,20) & 0.001 & 0.011 & 0.001 & 0.720 \\
  11 & (8,25) & 0.953 & 0.969 & 0.972 & 0.251 \\ 
  12 & (8,26) & 0.241 & 0.998 & 0.389 & 1.000 \\ 
  13 & (9,26) & 0.004 & 0.021 & 0.010 & 0.601 \\ 
  14 & (10,15) & 0.000 & 0.000 & 0.000 & 0.999 \\ 
  15 & (10,16) & 0.001 & 0.641 & 0.001 & 1.000 \\ 
  16 & (10,18) & 0.000 & 0.010 & 0.000 & 0.996 \\ 
  17 & (10,21) & 0.009 & 0.003 & 0.007 & 0.987 \\ 
  18 & (10,26) & 0.001 & 0.005 & 0.002 & 1.000 \\ 
  19 & (11,16) & 0.000 & 0.984 & 0.002 & 0.010 \\ 
  20 & (11,18) & 0.056 & 0.980 & 0.045 & 0.001 \\
  21 & (14,20) & 0.652 & 0.008 & 0.660 & 0.972 \\
  22 & (15,16) & 0.000 & 0.004 & 0.000 & 0.999 \\ 
  23 & (15,26) & 0.991 & 0.993 & 0.994 & 0.013  \\
  24 & (16,17)  & 0.000 & 0.055 & 0.000 & 0.963 \\ 
  25 & (16,25) & 0.012 & 0.037 & 0.007 & 0.785 \\ 
  26 & (16,26) & 0.000 & 0.995 & 0.007 & 1.000 \\ 
  27 & (17,22) & 0.000 & 0.000 & 0.000 & 0.741 \\ 
  28 & (17,25) & 0.000 & 0.000 & 0.000 & 0.751 \\ 
  29 & (18,26) & 0.362 & 0.021 & 0.508 & 0.062 \\
  30 & (20,24) & 0.001 & 0.000 & 0.003 & 0.972 \\ 
\caption{Posterior inclusion probabilities not included in all the four compared priors for the pCR case.}
\label{Hess_pCR_omitted}
\end{tabularx}


\newpage

\section*{Appendix A - FINCS algorithm}

FINCS is a serial procedure which utilises three types of moves: local, resampling and global. The local moves depend on updated estimates of the posterior edge inclusion probabilities. Resampling of one of the previously visited models is done in proportion to the current estimate of their posterior probabilities. The global moves allow us to explore those regions that would not be accessible in a reasonable number of local steps and with the help of the local moves, they try to address the multimodality of the problem at hand. Clearly, FINCS is not an MCMC scheme, but a hybrid algorithm designed to explore a collection of likely graphs.\\
As suggested by \citet{Scott2008}, for small-to-moderate-sized graphs, the convergence of FINCS is quite fast irrespective of the starting graph. The necessity of the global move becomes apparent when the true graph  has a lot of vertices. Then, a version of FINCS with only local moves and resampling steps would get trapped in the local hills, a behaviour which was also observed with the standard Metropolis-Hastings. Moreover, taking into account the enormity of the graph space to be explored, even a global variant of FINCS would depend on the starting graph. Here, the original authors have used an initial estimated graph based on conditional regressions which is the default setting in the FINCS algorithm. \citet{Scott2008} recommend a mixture of $80\%$ to $90\%$ local moves with the remainder used for global moves. Out of those local moves, $10\%$ to  $15\%$ should be dedicated to the resampling step. In all considered simulated and real data analyses, we have used the default setting as in \citet{Scott2008}, namely a global version of the FINCS algorithm with a resampling step every 10 iterations and a global move used every 50 iterations.

\noindent Given the data and some parameters do the following steps:

\begin{algorithm}[H]
\SetAlgoLined

\vspace{0.2cm} 
 
\begin{itemize}
\item[\textbf{Step 1}] Initialise a graph based on edges suggested by conditional regressions, followed by triangulation.
\item[\textbf{Step 2}] Loop over the iterations in a serial manner:
	\begin{itemize}
	\item[1] At a certain number of iterations do a global move through a randomized median triangulation pair (see below for a definition). Starting from a random median graph, we add or delete an edge such that decomposability is maintained and the log score is improved
	\item[2] At a certain number of iteration we resample  one of the previous saved local graphs
	\item[3] Do a local move by deleting or adding an edge that maintains decomposability. When an edge is added, it is done in proportion to the estimated posterior probability of inclusion $\hat{q}_{ij}$ for edge $(i,j)$, whereas when there is a deletion, the edge is affected in inverse proportion to the estimated inclusion probabilities
\item[4] Save the local graph in a finite resampling list and remove those graphs that do not improve the log score.
	\end{itemize}
\end{itemize}
\end{algorithm}

According to \citet{Scott2008}, a \textit{randomized median triangulation pair} represents a pair of decomposable graphs chosen in a certain way from the median graph $G_N$ which will often be non-decomposable. One of the pair members will be the minimal decomposable supergraph  $G^{+}\supset G_N$, whilst the other will be the maximal decomposable subgraph $G^{-}\subset G_N$. Based on the posterior probabilities, we choose one of $G^+$ or $G^-$ as our current generated graph at the respective iteration step. This randomised median triangulation pair allows the exploration of new regions in the decomposable graph space.

\section*{Appendix B - Loss in information associated to the full graph}

In this appendix we study the loss in information associated to the full graph. In particular, we show that the expected Kullback--Leibler divergence for the complete graph, although strictly positive, is closed to zero and, the larger the graph space the smaller the divergence. This motivates the approximation to zero of the loss in information and therefore justifies the form of the prior as in \eqref{RefinedOurPrior}. The procedure we employed to compute the expected Kullback--Leibler divergence for the complete graph is as in  \citet{Jog}. We have considered graph size $|V|=3,5,10,50$ and, as prior on the covariance matrix, we have chosen a non-informative prior, i.e.  $G$-Wishart$(3,\mathbb{I}_{|V|})$  \citep{Massam2017}, as well as a more general set-up where the identity matrix is replaced by a symmetric positive matrix, i.e. $G$-Wishart$(3,D)$. The matrix $D$ has been chosen so that the elements of the main diagonal are equal $|V|$, while the off-diagonal elements are $|V|-1$. Note that, the effect of $D$ is to make the off-diagonal elements of the $G$-Wishart prior smaller than the case when $\mathbb{I}_{|V|}$ is used. 

\begin{figure}[h!]
  \centering
    \subcaptionbox{}{\includegraphics[width=0.6\textwidth]{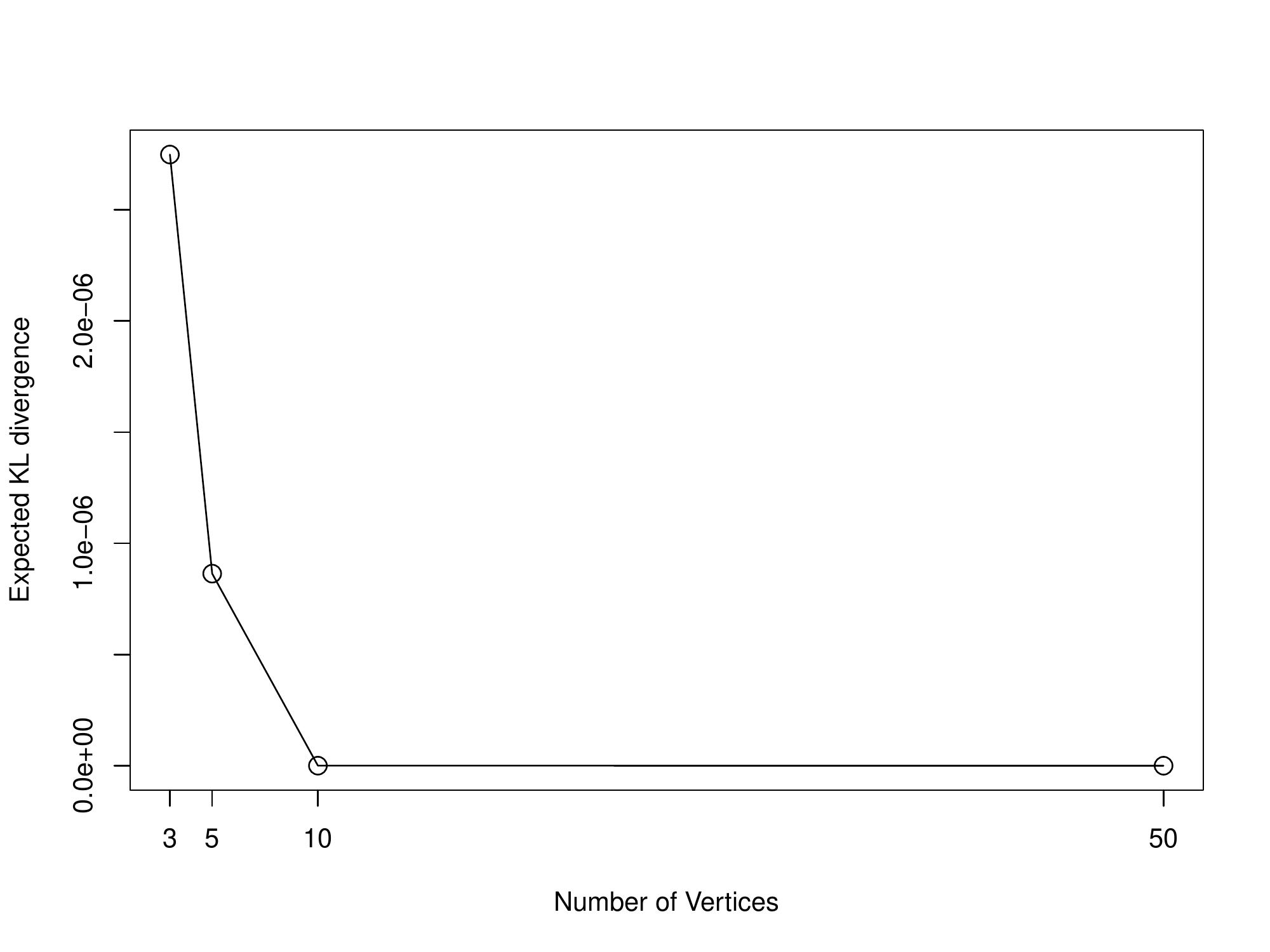}}
  \subcaptionbox{}{\includegraphics[width=0.6\textwidth]{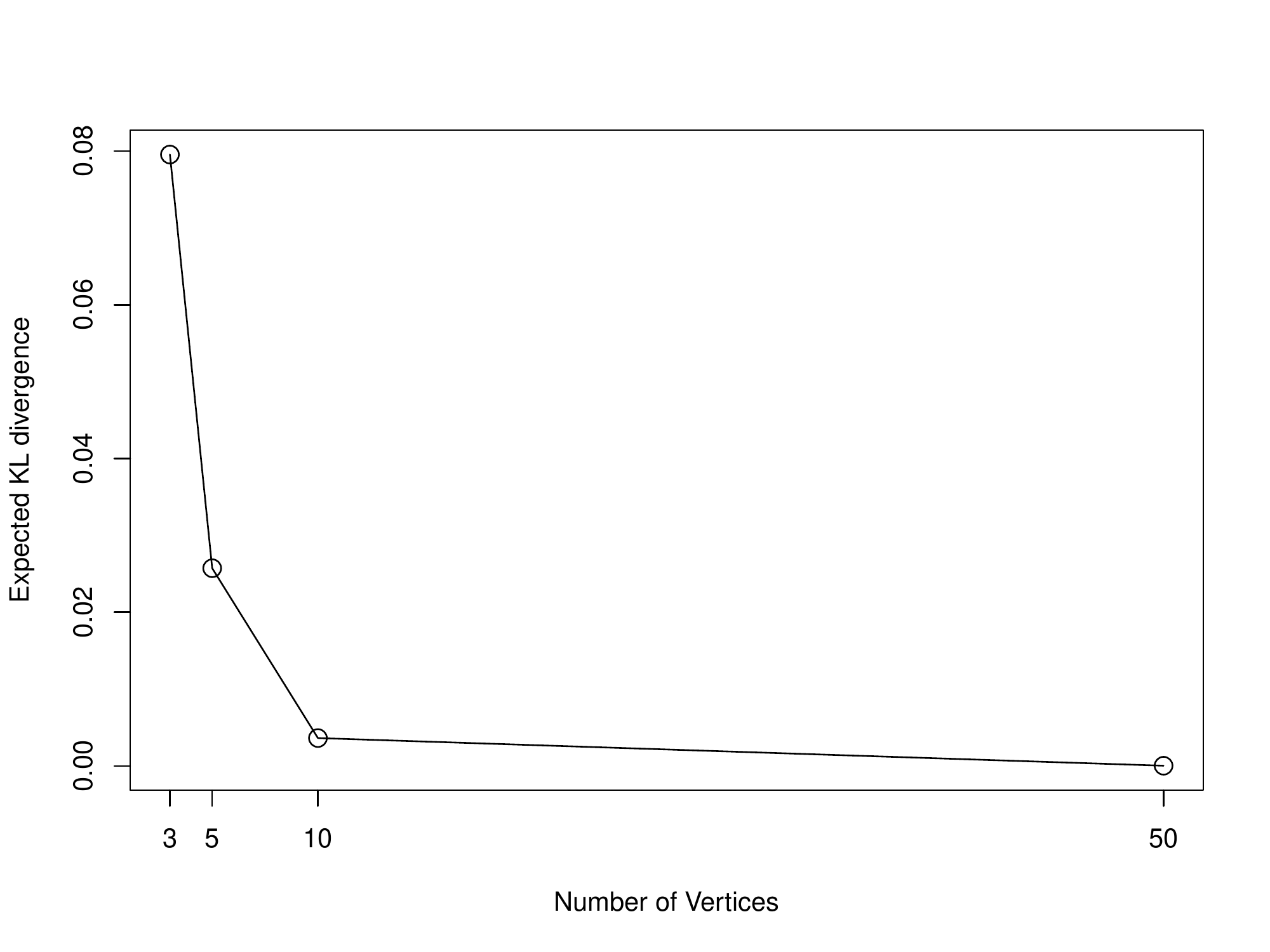}}\hfill
   \subcaptionbox{}{\includegraphics[width=0.6\textwidth]{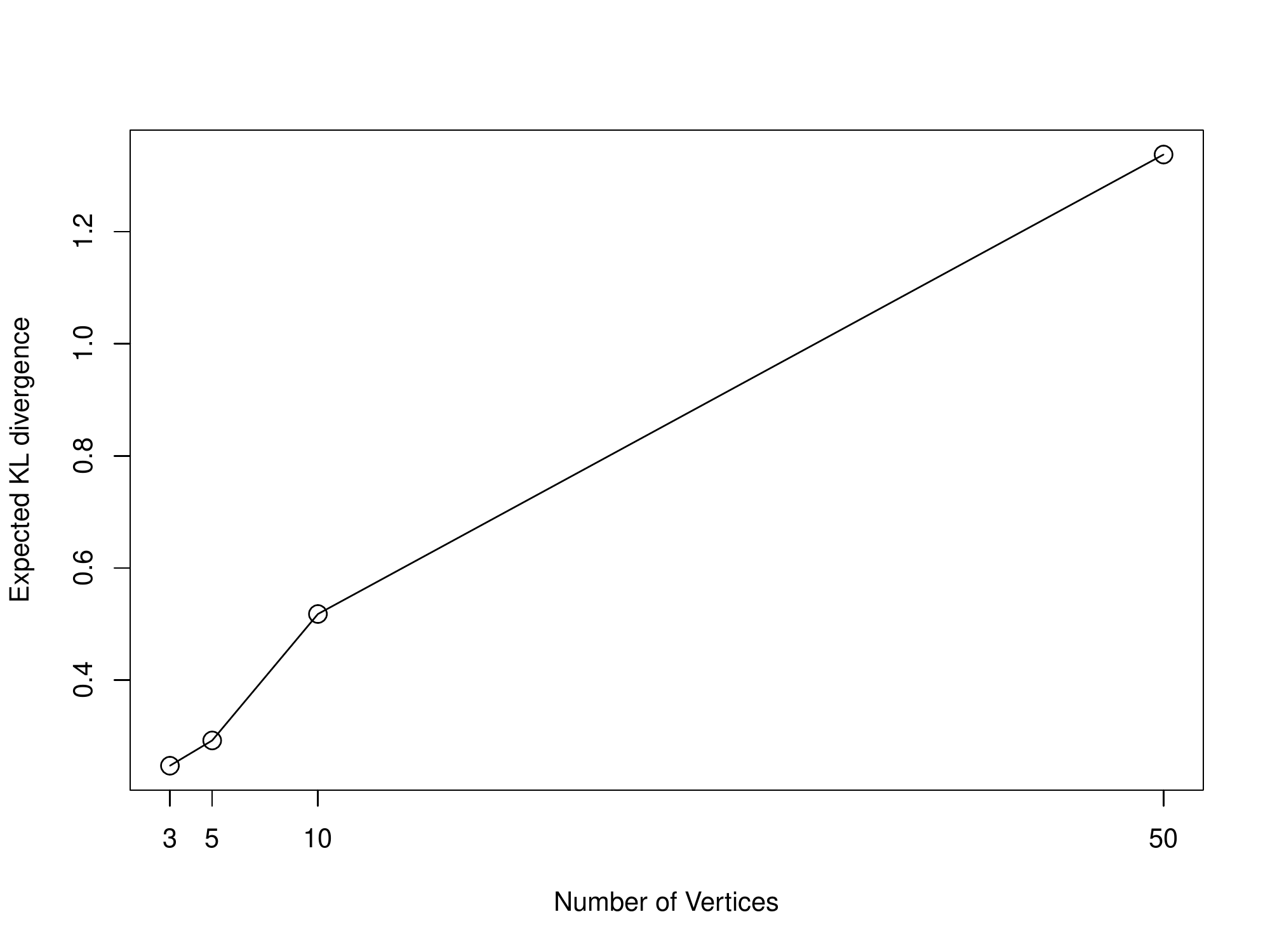}}
     \caption{Expected Kullback--Leibler divergence for the complete graph for different graph sizes. We have considered three different priors on the covariance matrix: (a) $G$-Wishart$(3,\mathbb{I}_{|V|})$, (b) $G$-Wishart$(3,D)$ and (c) $G$-Wishart$(3,D^{-1})$.}
    \label{fig:KL_Expected}
 \end{figure}

In Figure \ref{fig:KL_Expected}, we have plotted the expected Kullback--Leibler divergence for both cases: in panel (a) the $G$-Wishart$(3,\mathbb{I}_{|V|})$ and in panel (b) the $G$-Wishart$(3,D)$. Although when the identity matrix is used as hyperprior yield much smaller expected Kullback--Leibler divergences than when the matrix $D$ is used, in both cases we see that the loss in information associated to the complete graph quickly drops as the graph size increases. The estimated expected Kullback--Leibler divergencies have been obtained through a Monte Carlo procedure with a sample size of 1000. We noted that the Monte Carlo error decreases as $|V|$ increases for both the identity matrix and the $D$ matrix; in particular, we move from an error of the order $10^{-3}$ for $|V|=3$ to the order $10^{-9}$ for $|V|=50$. To further understand the role played by matrix $D$, we have also considered the parameter of the $G$-Wishart equal to $D^{-1}$. Although the matrix is still positive definite, the effect on the expected Kullback--Leibler divergence is opposite to the previous case, that is that the loss in information increases as the graph size increase, see panel (c) in Figure \ref{fig:KL_Expected}.

%
%

\label{Bibliography}
\bibliographystyle{agsm}  

\begin{thebibliography}{51}
\providecommand{\natexlab}[1]{#1}
\providecommand{\url}[1]{\texttt{#1}}
\providecommand{\urlprefix}{URL }
\expandafter\ifx\csname urlstyle\endcsname\relax
  \providecommand{\doi}[1]{doi:\discretionary{}{}{}#1}\else
  \providecommand{\doi}{doi:\discretionary{}{}{}\begingroup
  \urlstyle{rm}\Url}\fi
\providecommand{\eprint}[2][]{\url{#2}}

\bibitem[{Ambroise, Chiquet \& Matias(2009)}]{Ambroise2009}
\textsc{Ambroise, C., Chiquet, J. \& Matias, C.} (2009).
\newblock {Inferring sparse Gaussian graphical models with latent structure}.
\newblock \textsl{Electronic Journal of Statistics} \textbf{3}, 205--238.

\bibitem[{Armstrong {et~al.}(2009)Armstrong, Carter, Wong \&
  Kohn}]{Armstrong2009}
\textsc{Armstrong, H., Carter, C.K., Wong, K.F.K. \& Kohn, R.} (2009).
\newblock {Bayesian Covariance Matrix Estimation using a Mixture of
  Decomposable Graphical Models}.
\newblock \textsl{Statistics and Computing} \textbf{19}, 303--316.

\bibitem[{Atay-Kayis \& Massam(2005)}]{Massam}
\textsc{Atay-Kayis, A. \& Massam, H.} (2005).
\newblock {A Monte Carlo Method for Computing the Marginal Likelihood in
  Nondecomposable Gaussian Graphical Models}.
\newblock \textsl{Biometrika} \textbf{92}, 317--335.

\bibitem[{Banerjee, El~Ghaoui \& d'Aspremont(2008)}]{Banerjee2008}
\textsc{Banerjee, O., El~Ghaoui, L. \& d'Aspremont, A.} (2008).
\newblock {Model Selection Through Sparse Maximum Likelihood Estimation for
  Multivariate Gaussian or Binary Data}.
\newblock \textsl{The Journal of Machine Learning Research} \textbf{9},
  485--516.

\bibitem[{Bell \& King(2007)}]{Bell2007}
\textsc{Bell, P. \& King, S.} (2007).
\newblock {Sparse Gaussian graphical models for speech recognition}.
\newblock In \textsl{{INTERSPEECH} 2007, 8th Annual Conference of the
  International Speech Communication Association, Antwerp, Belgium, August
  27-31, 2007}. pp. 2113--2116.

\bibitem[{Bien \& Tibshirani(2011)}]{Bien2011}
\textsc{Bien, J. \& Tibshirani, R.J.} (2011).
\newblock Sparse estimation of a covariance matrix.
\newblock \textsl{Biometrika} \textbf{98}, 807--820.

\bibitem[{Bilmes(2004)}]{Bilmes2004}
\textsc{Bilmes, J.A.} (2004).
\newblock {Graphical Models and Automatic Speech Recognition}.
\newblock In \textsl{Mathematical Foundations of Speech and Language
  Processing}, eds. M.~Johnson, S.P. Khudanpur, M.~Ostendorf \& R.~Rosenfeld.
  New York, NY: Springer New York, pp. 191--245.

\bibitem[{Carvalho \& Scott(2009)}]{Carvalho}
\textsc{Carvalho, C.M. \& Scott, J.G.} (2009).
\newblock {Objective Bayesian Model Selection in Gaussian Graphical Models}.
\newblock \textsl{Biometrika} \textbf{96}, 497--512.

\bibitem[{Chiquet {et~al.}(2009)Chiquet, Smith, Grasseau, Matias \&
  Ambroise}]{SIMONE}
\textsc{Chiquet, J., Smith, A., Grasseau, G., Matias, C. \& Ambroise, C.}
  (2009).
\newblock {SIMoNe: Statistical Inference for MOdular NEtworks}.
\newblock \textsl{Bioinformatics} \textbf{25}, 417--418.

\bibitem[{Consonni, La~Rocca \& Peluso(2017)}]{Consonni2017}
\textsc{Consonni, G., La~Rocca, L. \& Peluso, S.} (2017).
\newblock {Objective Bayes Covariate-Adjusted Sparse Graphical Model
  Selection}.
\newblock \textsl{Scandinavian Journal of Statistics} \textbf{44}, 741--764.

\bibitem[{Cover \& Thomas(2006)}]{Cover2006}
\textsc{Cover, T.M. \& Thomas, J.A.} (2006).
\newblock \textsl{{Elements of Information Theory},}.
\newblock John Wiley and Sons.

\bibitem[{Cowell {et~al.}(2007)Cowell, Dawid, Lauritzen \&
  Spiegelhalter}]{Cowell2007}
\textsc{Cowell, R.G., Dawid, A.P., Lauritzen, S.L. \& Spiegelhalter, D.J.}
  (2007).
\newblock \textsl{Probabilistic Networks and Expert Systems: Exact
  Computational Methods for Bayesian Networks}.
\newblock Springer Publishing Company, Incorporated, 1st edn.

\bibitem[{Dawid \& Lauritzen(1993)}]{Dawid1993}
\textsc{Dawid, A.P. \& Lauritzen, S.L.} (1993).
\newblock {Hyper Markov Laws in the Statistical Analysis of Decomposable
  Graphical Models}.
\newblock \textsl{The Annals of Statistics} \textbf{21}, {1272--1317}.

\bibitem[{Dobra {et~al.}(2004)Dobra, Hans, Jones, Nevins, Yao \&
  West}]{Dobra2004}
\textsc{Dobra, A., Hans, C., Jones, B., Nevins, J.R., Yao, G. \& West, M.}
  (2004).
\newblock {Sparse graphical models for exploring gene expression data}.
\newblock \textsl{Journal of Multivariate Analysis} \textbf{90}, 196 -- 212.

\bibitem[{Dobra, Lenkoski \& Rodriguez(2011)}]{DobraRJMCMC}
\textsc{Dobra, A., Lenkoski, A. \& Rodriguez, A.} (2011).
\newblock {Bayesian Inference for General Gaussian Graphical Models With
  Application to Multivariate Lattice Data}.
\newblock \textsl{Journal of the American Statistical Association}
  \textbf{106}, 1418--1433.

\bibitem[{Friedman, Hastie \& Tibshirani(2008)}]{Friedman2008}
\textsc{Friedman, J., Hastie, T. \& Tibshirani, R.} (2008).
\newblock Sparse inverse covariance estimation with the graphical lasso.
\newblock \textsl{Biostatistics} \textbf{9}, 432--441.

\bibitem[{Friedman {et~al.}(2000)Friedman, Linial, Nachman \&
  Pe'er}]{Friedman2000}
\textsc{Friedman, N., Linial, M., Nachman, I. \& Pe'er, D.} (2000).
\newblock {Using Bayesian Networks to Analyze Expression Data}.
\newblock \textsl{Journal of Computational Biology} \textbf{7}, 601--620.

\bibitem[{Geiger \& Heckerman(2002)}]{Geiger2002}
\textsc{Geiger, D. \& Heckerman, D.} (2002).
\newblock {Parameter priors for directed acyclic graphical models and the
  characterization of several probability distributions}.
\newblock \textsl{The Annals of Statistics} \textbf{30}, 1412--1440.

\bibitem[{Giudici \& Green(1999)}]{Giudici}
\textsc{Giudici, P. \& Green, P.} (1999).
\newblock {Decomposable Graphical Gaussian Model Determination}.
\newblock \textsl{Biometrika} \textbf{86}, 785--801.

\bibitem[{Giudici \& Spelta(2016)}]{Giudici2016}
\textsc{Giudici, P. \& Spelta, A.} (2016).
\newblock {Graphical Network Models for International Financial Flows}.
\newblock \textsl{Journal of Business \& Economic Statistics} \textbf{34},
  128--138.

\bibitem[{Green(1995)}]{Green1995}
\textsc{Green, P.J.} (1995).
\newblock Reversible jump markov chain monte carlo computation and bayesian
  model determination.
\newblock \textsl{Biometrika} \textbf{82}, 711--732.

\bibitem[{Hess {et~al.}(2006)Hess, Anderson, Symmans, Valero, Ibrahim, Mejia,
  Booser, Theriault, Buzdar, Dempsey, Rouzier, Sneige, Ross, Vidaurre,
  G{\'o}mez, Hortobagyi \& Pusztai}]{Hess2006}
\textsc{Hess, K.R., Anderson, K., Symmans, W.F., Valero, V., Ibrahim, N.,
  Mejia, J.A., Booser, D., Theriault, R.L., Buzdar, A.U., Dempsey, P.J.,
  Rouzier, R., Sneige, N., Ross, J.S., Vidaurre, T., G{\'o}mez, H.L.,
  Hortobagyi, G.N. \& Pusztai, L.} (2006).
\newblock {Pharmacogenomic Predictor of Sensitivity to Preoperative
  Chemotherapy With Paclitaxel and Fluorouracil, Doxorubicin, and
  Cyclophosphamide in Breast Cancer}.
\newblock \textsl{Journal of Clinical Oncology} \textbf{24}, 4236--4244.

\bibitem[{Hinoveanu, Leisen \& Villa(2019)}]{Hinoveanu2019}
\textsc{Hinoveanu, L.C., Leisen, F. \& Villa, C.} (2019).
\newblock {Bayesian loss-based approach to change point analysis}.
\newblock \textsl{Computational Statistics \& Data Analysis} \textbf{129}, 61
  -- 78.

\bibitem[{Jog \& Loh(2015)}]{Jog}
\textsc{Jog, V. \& Loh, P.L.} (2015).
\newblock {On Model Misspecification and KL Separation for Gaussian Graphical
  Models}.
\newblock In \textsl{2015 IEEE International Symposium on Information Theory
  (ISIT)}. pp. 1174--1178.

\bibitem[{Jones {et~al.}(2005)Jones, Carvalho, Dobra, Hans, Carter \&
  West}]{jones2005}
\textsc{Jones, B., Carvalho, C., Dobra, A., Hans, C., Carter, C. \& West, M.}
  (2005).
\newblock {Experiments in Stochastic Computation for High-Dimensional Graphical
  Models}.
\newblock \textsl{Statistical Science} \textbf{20}, 388--400.

\bibitem[{Kundu, Mallick \& Baladandayuthapani(2019)}]{Kundu2019}
\textsc{Kundu, S., Mallick, B.K. \& Baladandayuthapani, V.} (2019).
\newblock {Efficient Bayesian Regularization for Graphical Model Selection}.
\newblock \textsl{Bayesian Analysis} \textbf{14}, 449--476.

\bibitem[{Lauritzen(1996)}]{Lauritzen:1996}
\textsc{Lauritzen, S.L.} (1996).
\newblock \textsl{Graphical Models}.
\newblock Clarendon Press, Oxford.

\bibitem[{Mair(2015)}]{Mair2015}
\textsc{Mair, P.} (2015).
\newblock \textsl{{APR: Applied Psychometrics With R}}.
\newblock R package version 0.0-6/r205.

\bibitem[{McNally {et~al.}(2015)McNally, Robinaugh, Wu, Wang, Deserno \&
  Borsboom}]{McNally2015}
\textsc{McNally, R.J., Robinaugh, D.J., Wu, G.W.Y., Wang, L., Deserno, M.K. \&
  Borsboom, D.} (2015).
\newblock {Mental Disorders as Causal Systems: A Network Approach to
  Posttraumatic Stress Disorder}.
\newblock \textsl{Clinical Psychological Science} \textbf{3}, 836--849.

\bibitem[{Meinshausen \& B{\"u}hlmann(2006)}]{Meinhausen2006}
\textsc{Meinshausen, N. \& B{\"u}hlmann, P.} (2006).
\newblock {High-Dimensional Graphs and Variable Selection with the Lasso}.
\newblock \textsl{The Annals of Statistics} \textbf{34}, 1436--1462.

\bibitem[{Merhav \& Feder(1998)}]{Merhav1998}
\textsc{Merhav, N. \& Feder, M.} (1998).
\newblock Universal prediction.
\newblock \textsl{IEEE Transactions on Information Theory} \textbf{44},
  2124--2147.

\bibitem[{Mohammadi {et~al.}(2017)Mohammadi, Abegaz, van~den Heuvel \&
  Wit}]{MohammadiCopula2017}
\textsc{Mohammadi, A., Abegaz, F., van~den Heuvel, E. \& Wit, E.C.} (2017).
\newblock {Bayesian modelling of Dupuytren disease by using Gaussian copula
  graphical models}.
\newblock \textsl{Journal of the Royal Statistical Society: Series C (Applied
  Statistics)} \textbf{66}, 629--645.

\bibitem[{{Mohammadi}, {Massam} \& {Letac}(2017)}]{Massam2017}
\textsc{{Mohammadi}, A., {Massam}, H. \& {Letac}, G.} (2017).
\newblock {The Ratio of Normalizing Constants for Bayesian Graphical Gaussian
  Model Selection}.
\newblock \textsl{ArXiv e-prints} \eprint{1706.04416}.

\bibitem[{Mohammadi \& Wit(2015)}]{Mohammadi2015}
\textsc{Mohammadi, A. \& Wit, E.C.} (2015).
\newblock {Bayesian Structure Learning in Sparse Gaussian Graphical Models}.
\newblock \textsl{{Bayesian Analysis}} \textbf{10}, 109--138.

\bibitem[{Mohammadi \& Wit(2017)}]{Moham}
\textsc{Mohammadi, A. \& Wit, E.C.} (2017).
\newblock \textsl{{BDgraph: Bayesian Structure Learning in Graphical Models
  using Birth-Death MCMC}}.
\newblock R package version 2.36.

\bibitem[{Newton {et~al.}(2004)Newton, Noueiry, Sarkar \&
  Ahlquist}]{Newton2004}
\textsc{Newton, M.A., Noueiry, A., Sarkar, D. \& Ahlquist, P.} (2004).
\newblock {Detecting differential gene expression with a semiparametric
  hierarchical mixture method}.
\newblock \textsl{Biostatistics} \textbf{5}, 155--176.

\bibitem[{O'Hagan(1995)}]{OHagan1995}
\textsc{O'Hagan, A.} (1995).
\newblock Fractional bayes factors for model comparison.
\newblock \textsl{Journal of the Royal Statistical Society. Series B
  (Methodological)} \textbf{57}, 99--138.

\bibitem[{Roverato \& Whittaker(1998)}]{Roverato1998}
\textsc{Roverato, A. \& Whittaker, J.} (1998).
\newblock {The Isserlis matrix and its application to non-decomposable
  graphical Gaussian models}.
\newblock \textsl{Biometrika} \textbf{85}, 711--725.

\bibitem[{Sachs {et~al.}(2005)Sachs, Perez, Pe{\textquoteright}er,
  Lauffenburger \& Nolan}]{Sachs2005}
\textsc{Sachs, K., Perez, O., Pe{\textquoteright}er, D., Lauffenburger, D.A. \&
  Nolan, G.P.} (2005).
\newblock {Causal Protein-Signaling Networks Derived from Multiparameter
  Single-Cell Data}.
\newblock \textsl{Science} \textbf{308}, 523--529.

\bibitem[{Scott \& Carvalho(2008)}]{Scott2008}
\textsc{Scott, J.G. \& Carvalho, C.M.} (2008).
\newblock Feature-inclusion stochastic search for gaussian graphical models.
\newblock \textsl{Journal of Computational and Graphical Statistics}
  \textbf{17}, 790--808.

\bibitem[{Shojaie \& Michailidis(2010)}]{Shojaie2010}
\textsc{Shojaie, A. \& Michailidis, G.} (2010).
\newblock {Penalized Principal Component Regression on Graphs for Analysis of
  Subnetworks}.
\newblock In \textsl{Advances in Neural Information Processing Systems 23},
  eds. J.D. Lafferty, C.K.I. Williams, J.~Shawe-Taylor, R.S. Zemel \&
  A.~Culotta. Curran Associates, Inc., pp. 2155--2163.

\bibitem[{Spirtes, Glymour \& Scheines(2000)}]{Spirtes2000}
\textsc{Spirtes, P., Glymour, C. \& Scheines, R.} (2000).
\newblock \textsl{Causation, Prediction, and Search}.
\newblock MIT press, 2nd edn.

\bibitem[{Stingo \& Marchetti(2015)}]{Stingo2015}
\textsc{Stingo, F. \& Marchetti, G.M.} (2015).
\newblock {Efficient local updates for undirected graphical models}.
\newblock \textsl{Statistics and Computing} \textbf{25}, 159--171.

\bibitem[{Stingo {et~al.}(2010)Stingo, Chen, Vannucci, Barrier \&
  Mirkes}]{Stingo2010}
\textsc{Stingo, F.C., Chen, Y.A., Vannucci, M., Barrier, M. \& Mirkes, P.E.}
  (2010).
\newblock {A Bayesian graphical modeling approach to microRNA regulatory
  network inference}.
\newblock \textsl{The Annals of Applied Statistics} \textbf{4}, 2024--2048.

\bibitem[{Tibshirani(1996)}]{Tibshirani1996}
\textsc{Tibshirani, R.} (1996).
\newblock Regression shrinkage and selection via the lasso.
\newblock \textsl{Journal of the Royal Statistical Society. Series B
  (Methodological)} \textbf{58}, 267--288.

\bibitem[{{Villa} \& {Lee}(2020)}]{VillaLee}
\textsc{{Villa}, C. \& {Lee}, J.E.} (2020).
\newblock {A Loss-Based Prior for Variable Selection in Linear Regression
  Methods}.
\newblock \textsl{Bayesian Analysis} \textbf{15}, 533--558.

\bibitem[{Villa \& Walker(2015)}]{VillaModel}
\textsc{Villa, C. \& Walker, S.} (2015).
\newblock {An Objective Bayesian Criterion to Determine Model Prior
  Probabilities}.
\newblock \textsl{Scandinavian Journal of Statistics} \textbf{42}, 947--966.

\bibitem[{Wang {et~al.}(2016)Wang, Ren, Ding, Fang, Sun, MacDonald, Sweet, Wang
  \& Chen}]{Wang2016}
\textsc{Wang, T., Ren, Z., Ding, Y., Fang, Z., Sun, Z., MacDonald, M.L., Sweet,
  R.A., Wang, J. \& Chen, W.} (2016).
\newblock {FastGGM: An Efficient Algorithm for the Inference of Gaussian
  Graphical Model in Biological Networks}.
\newblock \textsl{PLOS Computational Biology} \textbf{12}, 1--16.

\bibitem[{Williams(2018)}]{Williams2018}
\textsc{Williams, D.R.} (2018).
\newblock Bayesian inference for gaussian graphical models: Structure learning,
  explanation, and prediction.
\newblock \textsl{PsyArXiv} \eprint{x8dpr}.

\bibitem[{Yajima {et~al.}(2015)Yajima, Telesca, Ji \& Müller}]{Yajima2015}
\textsc{Yajima, M., Telesca, D., Ji, Y. \& Müller, P.} (2015).
\newblock {Detecting differential patterns of interaction in molecular
  pathways}.
\newblock \textsl{Biostatistics} \textbf{16}, 240--251.

\bibitem[{Yuan \& Lin(2007)}]{Yuan}
\textsc{Yuan, M. \& Lin, Y.} (2007).
\newblock {Model Selection and Estimation in the Gaussian Graphical Model}.
\newblock \textsl{Biometrika} \textbf{94}, 19--35.

\end{thebibliography}

\end{document}